\documentclass[aps,prapplied,superscriptaddress,footinbib,floatfix,twocolumn]{revtex4-2}
\usepackage{graphicx}
\usepackage{multirow}
\usepackage{color}
\usepackage{amsmath}
\usepackage{amsfonts}
\usepackage{amssymb}
\usepackage{graphicx}
\graphicspath{{./}}
\usepackage{epstopdf}
\usepackage{empheq}
\usepackage{float}
\usepackage{cases}
\usepackage{mathtools}
\usepackage{dcolumn}
\usepackage{bbold}
\usepackage{bm}
\usepackage{ulem}
\usepackage[T1]{fontenc}

\usepackage[dvipsnames]{xcolor}
\usepackage[colorlinks]{hyperref}
\hypersetup{
colorlinks=True,
linkbordercolor=White,
linkcolor=BrickRed,
citecolor=Blue,	 
}

\begin{document}

\title{Variability of hole spin qubits in planar Germanium}

\author{Biel Martinez}
\email{biel.martinezidiaz@cea.fr}
\affiliation{Univ. Grenoble Alpes, CEA, LETI, F-38000, Grenoble, France}
\author{Yann-Michel Niquet}
\email{yniquet@cea.fr}
\affiliation{Univ. Grenoble Alpes, CEA, IRIG-MEM-L\_Sim, F-38000, Grenoble, France}

\date{\today}

\begin{abstract}

Hole spin qubits in Ge/GeSi heterostructures benefit from the clean environment of epitaxial interfaces and from the intrinsic spin-orbit coupling that enables efficient electrical control, which makes them promising candidates for quantum computation. However, spin-orbit coupling also enhances the sensitivity to electrical disorder, potentially increasing variability. In this work, we perform numerical simulations in a realistic device geometry to quantify the variability of the charge and spin properties of Ge qubits induced by charge traps at the SiGe/oxide interfaces. We show that while the variability of charge properties remains moderate, spin properties ($g$-factors and Rabi frequencies) show significant dispersion. We explore the implications of this variability for large-scale architectures, and provide guidelines to minimize variability both in terms of interface quality requirements and optimal operation strategies.

\end{abstract}

\maketitle

\section{Introduction}

Semiconductor spin qubits hold great promise for addressing the challenges of large-scale quantum computers \cite{Loss1998,Hanson2007}. The electrical confinement of electrons and holes into quantum dots (QD) of nanometric size, and the use of their spin as a coherent carrier of information has been successfully demonstrated in a variety of materials including GaAs/AlGaAs \cite{Petta2005,Mortemousque2021}, Si/SiO2 \cite{Maurand2016, Geyer2022,Klemt2023, Yang2020}, Si/SiGe \cite{Philips2022,Yoneda2018,koch2024}, and Ge/SiGe \cite{Hendrickx2020,Jirovec2021,Hendrickx2021,Hendrickx2023,john2024}. Interest has gradually focused on the last three materials because they can be isotopically purified from nuclear spins, thereby enhancing spin coherence \cite{Struck2020,Unseld2023,Stano2021,cvitkovich2024}. In the past few years, high-fidelity single- \cite{Yoneda2018,Xue2022,steinacker2024,Huang2024} and two-qubit gates \cite{steinacker2024,Huang2024,Xue2022} have been demonstrated in these platforms, as well as spin shuttling \cite{Noiri2022,Seidler2022,vanRiggelen-Doelman2024} and spin-photon coupling \cite{Samkharadze2018,yu2022strong,dijkema2023}, the basic ingredients for quantum links between two-dimensional arrays \cite{Mortemousque2021,Hendrickx2021,Borsoi2023,Unseld2023,HanLim2025}. This completes the toolkit needed to build large-scale spin qubit architectures \cite{Vandersypen2017,Vinet2018,Künne2024}. 

Metal-oxide-semiconductor (MOS) spin qubits naturally benefit from the miniaturization and fabrication capabilities of classical microelectronics technologies \cite{DeFranceschi2016, Vinet2018, Vinet2021}. However, the proximity with possibly charged amorphous materials and the defective nature of the Si/SiO$_2$ interface has been identified as a major source of qubit-to-qubit variability that could compromise scalability \cite{Martinez2022, Cifuentes2023}. In this context, epitaxial heterostructure devices have emerged as an alternative platform with a cleaner environment for the qubits, since the amorphous materials and defective interfaces are shifted a few tens of nanometers above the active layer \cite{Sammak2019,Scappucci2020,Scappucci2021,DegliEsposti2024}. Remarkable progress has been made recently in these heterostructures, with the demonstration of increasingly larger arrays of qubits \cite{Hendrickx2020,Hendrickx2021,Takeda2021,Philips2022,Wang2024,john2024,tosato2025}. In particular, hole spin qubits in Ge/GeSi heterostructures have become a compelling platform for quantum computing and simulation \cite{Wang2023,Zhang2025}.

The intrinsic spin-orbit coupling (SOC) in hole spin qubits indeed enables the electrical manipulation of spins with radio-frequency electric fields \cite{Golovach2006} or spin shuttling \cite{Wang2024} without the need for extrinsic elements such as micro-magnets. Nonetheless, SOC is a double-edge sword, as coupling the spins to electric fields opens the door to disorder-induced variability and extra dephasing channels. As for dephasing, the existence of sweet spots with enhanced coherence has been widely reported in hole spin qubits \cite{Piot2022,Hendrickx2023,Mauro2024,carballido2024,bassi2024}. Qubit-to-qubit variability, however, is a critical roadblock for holes in Si MOS devices \cite{Martinez2022}. While it does not represent a major limitation at present for few qubits in Ge/SiGe heterostructures, whether it can hinder scalability remains an open question.

In this work, we quantify the variability of the charge and spin properties of hole spin qubits in Ge/GeSi heterostructures by means of numerical simulations on realistic, disordered devices. We introduce the devices and methodology in Section \ref{sec:methodology}, then quantify the variability of the charge and spin properties of the qubits in Section \ref{sec:results}, and finally discuss the implications for the operation of large-scale quantum processors in Section \ref{sec:implications}. 

\section{Methodology} \label{sec:methodology}

\begin{figure}[t]
\centering
\includegraphics[width=0.98\columnwidth]{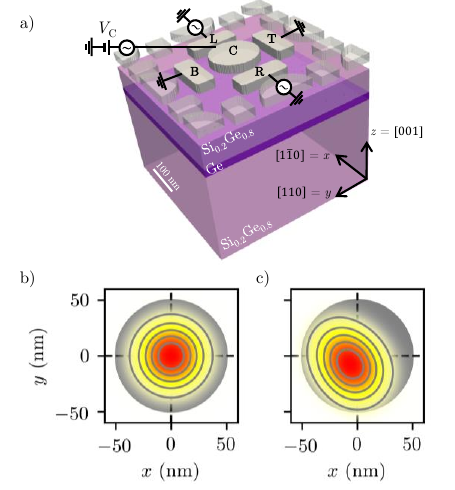}
\caption{a) Schematic representation of the simulated device. The active Ge layer is shown in purple, the top and bottom SiGe layers in pink, and the metallic gates in grey. The embedding oxide has been removed for clarity. b, c) Cross section of the ground-state density in the ($xy$) plane at the center of the Ge well for b) the pristine device and c) a defective device with $V_{\rm C}=-57.4$\,mV and $n_i=10^{11}$\,cm$^{-2}$. The shaded grey area highlights the position and dimensions of the C gate above.}
\label{fig:device}
\end{figure}

We focus this variability study on the impact of charge traps on the charge and spin properties of Ge/SiGe QDs, for a prototypical device similar to Refs. \cite{Martinez2022,Martinez2022-inhom,Abadillo2022,Mauro2024} (see Fig. \ref{fig:device}a). This device emulates a unit cell of the two-dimensional array architecture of Refs. \cite{Wang2024,john2024}. We compute the variability induced by charge traps at the semiconductor/oxide interface. Such defects are believed to be the dominant source of variability in Si MOS spin qubits \cite{Martinez2022}, and the cause of sizable energy fluctuations in Ge hole spin qubits \cite{Martinez2024}. We neglect other sources of variability such as dislocations or fabrication imperfections. We discuss the impact of residual long-range surface roughness \cite{peña2023} at the Ge/SiGe interfaces in Appendix \ref{app:SR}.

The simulated device is made of a 16 nm thick Ge well grown on a thick Ge$_{0.8}$Si$_{0.2}$ buffer and covered with a 50 nm thick Ge$_{0.8}$Si$_{0.2}$ barrier. In each unit cell, a central gate C with a diameter of 100 nm confines the holes in a QD; and a set of four lateral gates (labelled L, R, T, B) controls the interactions between the QDs. They are all grounded throughout this study \footnote{Note that any rigid  shift of all gate voltages leaves the electric field and confinement invariant.}, yet used to apply RF signals to drive the spin. The gates are 20 nm thick and are arranged in two layers: the L, R, T, B gates in a first layer on top of a 7 nm thick oxide (Al$_2$O$_3$), and the C gate in a second layer on top of a 14 nm thick oxide. The whole structure is encapsulated in 7 nm of Al$_2$O$_3$. The Ge well experiences a biaxial strain $\varepsilon_{xx}=\varepsilon_{yy}=\varepsilon_\parallel=-0.61\%$ and $\varepsilon_{zz}=\varepsilon_\perp=+0.45\%$ as a result of the lattice mismatch between Si and Ge, along with a residual in-plane strain $\varepsilon_{xx}=\varepsilon_{yy}=+0.26\%$ in the buffer \cite{Sammak2019}. Additionally, we include the inhomogeneous strains imprinted by the thermal contraction of the aluminium gates at cryogenic temperatures \cite{Abadillo2022,Reeber96,Corley-Wiciak2023}.

We randomly distribute point charges at the top Ge$_{0.8}$Si$_{0.2}$/Al$_2$O$_3$ interface to emulate dangling bonds of Si and Ge atoms, which we assume positively charged. We consider charge trap densities in the range $n_i=10^{10}-10^{12}$\,cm$^{-2}$. The upper bound is the typical estimate for SiGe/oxide interfaces, while the lower bound highlights the gains in improving interface quality \cite{massai2023}. Following the same computational workflow as in Refs. \cite{Martinez2022,Martinez2022-inhom,Abadillo2022,Mauro2024}, we solve Poisson's equation for the potential of the gates and trapped charges on a 3D mesh of the device, enforcing periodic boundary conditions along $x$ and $y$. The simulation box thus contains 2$\times$2 unit cells (with only the central dot occupied) to prevent significant interactions with the replicas of the disorder. The trapped charges are therefore screened by the dielectrics and metal gates. Then, we diagonalize a four-band Luttinger-Kohn Hamiltonian discretized with finite-differences on the same mesh. We obtain that way the single-hole ground-state energy ($E(1)=\mu$) and wave function in this potential (see Figs. \ref{fig:device}b and \ref{fig:device}c for an example). We also compute the charging energy as
\begin{equation} \label{eq:U}
U=E(2)-2E(1),
\end{equation}
using a full configuration interaction (CI) method \cite{Abadillo2021,rodriguez2025} for the two-particle ground-state energy $E(2)$. Finally, we make use of the $g$-matrix formalism \cite{Venitucci2018} to calculate the spin properties of the qubit, namely the $g$-factors, Rabi frequencies ($f_R$) and dephasing times ($T_2^*$).  

For each $n_i$, we collect statistics on 500 to 2000 simulations with independent, random realizations of the disorder. We apply the same gate voltage $V_{\rm C}$ for all realizations of a given $n_i$. We choose $V_{\rm C}$ so that the extensions of the QD wave function are $\ell_x=\ell_y=20$\,nm (with $\ell_u=\sqrt{\langle u^2\rangle-\langle u\rangle^2}$) in a ``pristine'' device with a uniform charge distribution $n_i$ that accounts for the average effect of the charge traps (equivalently the radius of the QD is $r_\parallel=\sqrt{\ell_x^2+\ell_y^2}=28.3$\,nm). The gate voltages are, therefore, $V_{\rm C}=-53.7$, $-54.3$, $-55.4$, $-57.4$, $-63.4$, $-73.9$, $-96.3$\,mV for $n_i=10^{10}$, $2.5 \times 10^{10}$, $5 \times10^{10}$, $10^{11}$, $2.5 \times10^{11}$, $5 \times10^{11}$, $10^{12}$\,cm$^{-2}$, respectively. Note that larger gate potentials are needed to reach the same QD size for increasing $n_i$, as the traps ultimately rule the electrostatics of the QDs. The charging energy $U$ of the pristine devices is $U\approx 2.4$\,meV whatever $n_i$.

We then compute, for each relevant property $f$, its median $\overline{f}$, the Inter Quartile Range (IQR, hereafter labeled $\mathcal{R}$) and the relative Inter Quartile Range ($\tilde{\mathcal{R}}$) from the percentiles ($P$) of the distribution as follows:
\begin{subequations}
\begin{align}
\overline{f}&=P(50\%), \\
\mathcal{R}(f)&=P(75\%)-P(25\%), \\ 
\tilde{\mathcal{R}}(f)&=\mathcal{R}(f)/\overline{f}\,.\label{eq:sigma}
\end{align}
\end{subequations}
We work with the median and IQR rather than the average and standard deviation as they are more robust and representative for non-normal distributions \footnote{Note that for a normal distribution, the median is equal to the mean, and the IQR is roughly 1.35 times the standard deviation.}. Moreover, we make use of a Bootstrap resampling method to assess the convergence of the statistics \cite{Davison1997}, and for each property we provide the 95\% confidence intervals as error bars. 

\section{Results and discussion} \label{sec:results}

In this Section we discuss the variability results. We focus on the variability of the charge properties in Subsection \ref{sub:charge}, and on the variability of the spin properties in Subsection \ref{sub:spin}. 

\subsection{Charge properties} \label{sub:charge}

To quantify the variability of the charge properties of Ge qubits, we monitor the chemical potential shift ($\Delta \mu$), the charging energy ($U$), the in-plane displacement ($d_\parallel$), and the in-plane size ($r_\parallel$) of the QDs. $\Delta \mu=\mu- \mu^0$ is the variation of the single-hole chemical potential $\mu$ induced by disorder, with $\mu^0$ the chemical potential of the pristine device. The charging energy $U$ is computed with CI, and the position and radius of the QD are defined as 
\begin{subequations}\label{eq:rpar}
\begin{align}
d_\parallel&=\sqrt{\langle x\rangle^2+\langle y\rangle ^2} \\
r_\parallel&=\sqrt{\langle x^2\rangle+\langle y^2\rangle-d_\parallel^2} ,
\end{align}
\end{subequations}
where $\langle x^n\rangle$ and $\langle y^n\rangle$ are the expectation values of $x^n$ and $y^n$ respectively.

\begin{figure}[t]
\centering
\includegraphics[width=0.98\columnwidth]{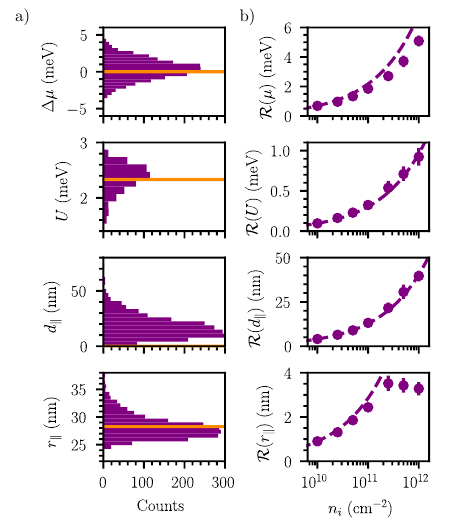}
\caption{Variability of the charge properties of Ge QDs in the device of Fig. \ref{fig:device}a. a) Histogram of the chemical potential shifts $\Delta \mu$, the charging energy $U$, the in-plane displacements $d_\parallel$, and the in-plane radius $r_\parallel$ of the QDs for a charge trap density $n_i=10^{11}$\,cm$^{-2}$ at $V_{\rm C}=-57.4$\,mV. The horizontal orange lines are the values for the pristine device. b) Dependence of the IQR of $\Delta \mu$, $U$, $d_\parallel$, and $r_\parallel$ on the charge trap density $n_i$. Dashed purple lines are guides to a $\propto\sqrt{n_i}$ dependence of the IQR.}
\label{fig:charge}
\end{figure}

In Fig. \ref{fig:charge}a we plot the histograms of the above charge properties for a trap density $n_i=10^{11}$\,cm$^{-2}$. $\Delta \mu$ shows fairly normal distributions, while $U$, $d_\parallel$ and $r_\parallel$ are slightly tailed (in part because they are positive quantities). The distribution of $U$'s mirrors the distribution of $r_\parallel$'s, as $U\propto 1/r_\parallel$ in a first approximation. All medians are in good agreement with the pristine device data at this $n_i$ (orange line), except for $d_\parallel$ (which can not be centered around zero because it is positive). The IQR of the distributions are plotted as a function of the trap density in Fig. \ref{fig:charge}b. The variability of the chemical potential reaches nearly 6 meV for $n_i=10^{12}$\,cm$^{-2}$. This is in line with the estimations of Ref. \cite{Martinez2024} for double Ge QDs, and significantly better than for holes in Si MOS. As discussed therein, fluctuations of the chemical potential can compromise the implementation of two-qubit gates, but remain manageable for Ge QDs. The absolute variability of the charging energies $U$ is much smaller ($\mathcal{R}(U)\approx 925$\,$\mu$eV for $n_i=10^{12}$\,cm$^{-2}$). The operation of large-scale quantum processors would certainly benefit from $\mathcal{R}(\mu)\ll U$, as disorder would not compromise the occupancy of the qubits. Note that this regime is only accessible at low trap densities $n_i\lesssim 5\times 10^{10}$\,cm$^{-2}$. We discuss the impact of these results on the reliability of crossbar architectures in Section \ref{sec:crossbar}.

The variations of the position and size of the disordered QDs remain moderate. The strong structural confinement along $z$ yields to negligible fluctuations in $\langle z\rangle$ and $\ell_z$ (not shown). The vast majority of the QDs are well located under the C gate, even though a few ones undergo sizable displacements (13\% of the devices move by $d_\parallel>30$\,nm at $n_i=10^{11}$\,cm$^{-2}$, the gate diameter being $d_{\rm C}=100$\,nm). The fluctuations of $r_\parallel$ are also small, with $\mathcal{R}(r_\parallel)<4$\,nm whatever $n_i$. We provide in Fig. \ref{fig:device}c a worst-case example of a deformed QD. These results show that charge disorder in the gate stack of Ge/SiGe heterostructures does not critically compromise the formation of the QDs, contrarily to what has been predicted for Si MOS devices \cite{Martinez2022}. This translates into a rather robust lever-arm $\alpha\approx 0.2$, with an IQR of only 0.04 for $n_i = 10^{12}$\,cm$^{-2}$. The lever arm is not, therefore, a relevant metric to assess QD homogeneity in this system, as it is rather insensitive to the moderate fluctuations in QD position and size. 

The relative IQR of all charge properties shows the approximate $\propto\sqrt{n_i}$ behavior expected from perturbation theory \cite{Martinez2022}, at least at low $n_i$. However, $\mathcal{R}(r_\parallel)$ strikingly saturates when $n_i>5 \times 10^{11}$\,cm$^{-2}$. We remind that the calculations are made at the bias $V_{\rm C}$ such that $r_\parallel=r_\parallel^0=28.3$\,nm in the pristine device with a uniform charge distribution at the SiGe/oxide interface. At low $n_i$, $\overline{r}_\parallel$ is approximately equal to $r_\parallel^0$, but when $n_i>5 \times 10^{11}$\,cm$^{-2}$, $\overline{r}_\parallel\lesssim r_\parallel^0$ (e.g., $\overline{r}_\parallel=23.8$\,nm at $n_i=10^{12}$\,cm$^{-2}$). The response to the stronger charge disorder indeed becomes non-linear and the median device is not the pristine device any more. The decrease of $\overline{r}_\parallel$ and saturation of $\mathcal{R}(r_\parallel)$ are fingerprints of these non-linearities, particularly prominent on these two quantities.

\subsection{Spin properties}\label{sub:spin}

We next discuss the variability of the spin properties. We address the $g$-factors in Sec. \ref{sec:gfact}, the Rabi frequencies in Sec. \ref{sec:Rabi}, and the dephasing times in Sec. \ref{sec:sweet}. 

\subsubsection{$g$-factors}\label{sec:gfact}

\begin{figure*}[t]
\centering
\includegraphics[width=0.98\textwidth]{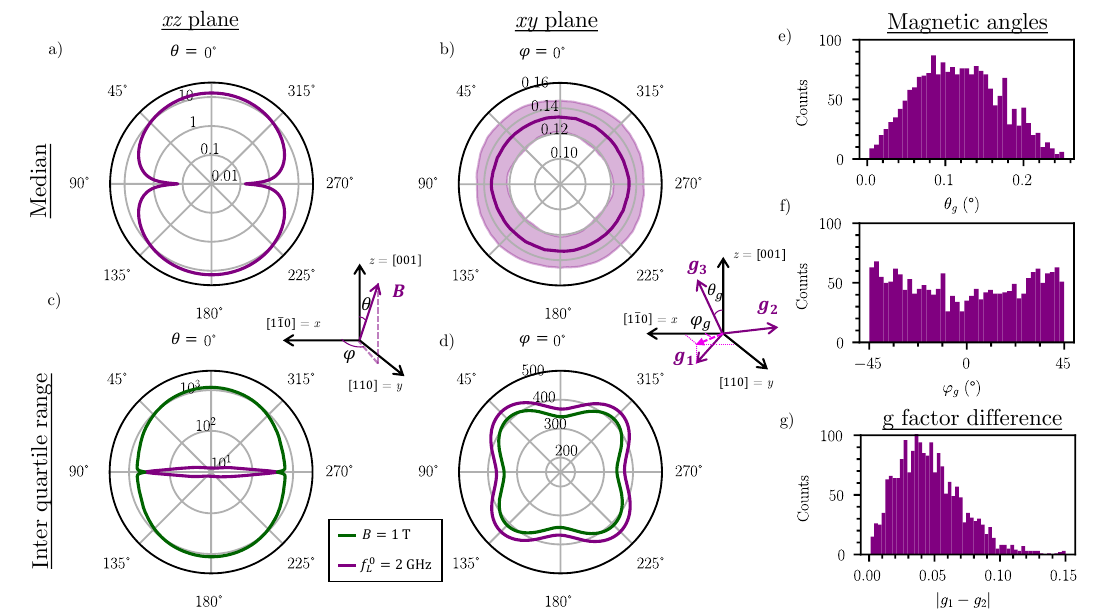}
\caption{Effective $g$-factor and magnetic axes variability for a charge trap density $n_i=10^{11}$\,cm$^{-2}$. a-b) Dependence of the effective $g$-factor on the angles $\theta$ and $\varphi$ of the magnetic field $\boldsymbol{B}=B(\sin(\theta)\cos(\varphi),\,\sin(\theta) \sin(\varphi),\,\cos(\theta))$. The shaded areas is the inter-quartile range. c-d) Dependence of the IQR $\mathcal{R}(f_L)$ (in MHz) on the magnetic field orientation. The results are plotted at constant $B=1$ T in green, and at the magnetic field amplitude such that $f_L=f_L^0=2$\,GHz in the pristine device in purple. e-f) Distribution of $\theta_g$ and $\varphi_g$. g) Distribution of the difference $|g_1-g_2|$ between the two principal in-plane $g$-factors. The insets define the angles $\theta$ and $\varphi$ that characterize the orientation of the magnetic field, and the angles $\theta_g$ and $\varphi_g$ that characterize the tilt of the magnetic axes. At constant $f_L$, the $g$-factor variability is maximal for in-plane magnetic fields. The in-plane magnetic angle $\varphi_g$ is actually almost homogeneously distributed.}
\label{fig:g_factors_1}
\end{figure*}

To calculate the Larmor frequencies $f_L=E_z/h$, we evaluate the $g$-matrix $\hat{g}$ of each qubit from the wave functions at zero magnetic field \cite{Venitucci2018}. The Zeeman splitting $E_z$ then reads
\begin{equation}
     E_z = \mu_B \sqrt{\boldsymbol{B}^T \hat{G} \boldsymbol{B}}= \mu_B  g^*|\boldsymbol{B}|\,,
\end{equation}
where $\mu_B$ is the Bohr magneton, $\boldsymbol{B}$ is the magnetic field, 
\begin{equation}
    g^*=\sqrt{\boldsymbol{B}^T \hat{G} \boldsymbol{B}}/|\boldsymbol{B}|
\end{equation}
is the effective $g$-factor, and
\begin{equation}
    \hat{G} = \hat{g}^T\hat{g}
\end{equation}
is the symmetric Zeeman tensor, which is an experimentally measurable quantity \cite{Crippa2018}. The eigenvalues of $\hat{G}$ are the squares of the principal $g$-factors $g_1$, $g_2$ and $g_3$, and the eigenvectors of $\hat{G}$ are the principal magnetic axes, which may not align with the device axes (the crystallographic $x$, $y$, $z$ directions). We quantify this misalignment by two angles $\theta_g$ and $\varphi_g$. $\theta_g$ is the polar angle of the magnetic axis closest to $z$ (the tilt of this magnetic axis with respect to $z$), while $\varphi_g$ is the azimuthal angle of the magnetic axis closest to $x$ (the angle between $x$ and the in-plane projection of this magnetic axis, see Fig. \ref{fig:g_factors_1}).

We plot in Fig. \ref{fig:g_factors_1} the median and IQR of the effective $g$-factor as a function of the magnetic field orientation, as well as the distribution of the angles $\theta_g$ and $\varphi_g$ ($n_i=10^{11}$\,cm$^{-2}$). We also plot the distribution of the absolute difference $|g_1-g_2|$ between the two principal ``in-plane'' $g$-factors (namely, those whose corresponding magnetic axes are closest to the plane).

As a consequence of disorder, the individual qubits may show anisotropic in-plane $g$-factors (Fig. \ref{fig:g_factors_1}g). Nevertheless, the median effective $g$-factor has the cylindrical symmetry. It matches the effective $g$-factor of the pristine device, reaching about 13.7 for out-of-plane and 0.13 for in-plane magnetic fields for all $n_i$ (Figs. \ref{fig:g_factors_1}a-b). The dispersion of the $g$-factors, highlighted by the shaded area, results in the detuning of the Larmor frequencies of the qubits. We thus plot $\mathcal{R}(f_L)$ in Figs. \ref{fig:g_factors_1}c-d, at constant magnetic field amplitude $B=1$\,T. $\mathcal{R}(f_L)$ shows a strong dependence on $\theta$ owing to the anisotropy of the $g$-factors. The absolute variability is maximal for $\boldsymbol{B}\parallel\boldsymbol{z}$ [$\mathcal{R}(f_L,\boldsymbol{B}\parallel\boldsymbol{z})=1172$\,MHz/T while $\mathcal{R}(f_L,\boldsymbol{B}\parallel\boldsymbol{x})=344$\,MHz/T]. However, the relative variability $\tilde{\mathcal{R}}(f_L)=\mathcal{R}(f_L)/\overline{f}_L$ is much larger in plane [$\tilde{\mathcal{R}}(f_L,\boldsymbol{B}\parallel\boldsymbol{z})=6$\,MHz/GHz and $\tilde{\mathcal{R}}(f_L,\boldsymbol{B}\parallel\boldsymbol{x})=186$\,MHz/GHz]. This quantity actually characterizes the dispersion of the Larmor frequency when the amplitude of the magnetic field is adjusted so that the median Larmor frequency $\overline{f}_L\approx f_L^0$ is independent on the magnetic field orientation (with $f_L^0$ the Larmor frequency of the pristine device). It is, therefore, more relevant experimentally (as devices are usually tuned to operate at a target $f_L$). The IQR of the Larmor frequency at constant $f_L^0=2$\,GHz is plotted in Figs. \ref{fig:g_factors_1}c and \ref{fig:g_factors_1}d. Therefore, in real \textit{operando} conditions the variability of $f_L$ is maximal for in-plane magnetic fields. 

We next analyze the distribution of the angles $\theta_g$ and $\varphi_g$. Fig. \ref{fig:g_factors_1}e shows that one of the principal magnetic axes remains very close the growth direction $z=[001]$, $\overline{\theta}_g$ barely reaching $0.11^\circ$ at $n_i=10^{11}$\,cm$^{-2}$ \footnote{The IQR of $\theta_g$ is $0.08^\circ$ at $n_i=10^{11}$\,cm$^{-2}$. It reaches $\mathcal{R}(\theta_g)=0.09$ degrees for $n_i=10^{12}$\,cm$^{-2}$, and it goes down to $0.03^\circ$ for $n_i=10^{10}$\,cm$^{-2}$. Note that $\overline{\theta}_g \to 0$ when $n_i\to 0$.}. This small tilt essentially results from the inhomogeneous shear strains $\varepsilon_{xz}$ and $\varepsilon_{yz}$ imprinted by the differential contraction of the materials when the device is cooled down \cite{Abadillo2022}. When $d_\parallel$ is non-zero, the quantum dot experiences finite $\langle\varepsilon_{xz}\rangle$ and $\langle\varepsilon_{yz}\rangle$, which give rise to non-zero $g$-matrix elements $g_{xz}$ and $g_{yz}$. The angle $\varphi_g$ is much more distributed, which means that the in-plane axes are largely randomized by the disorder (see Fig. \ref{fig:g_factors_1}f). There are, nonetheless, two preferential orientations, $\varphi=45^\circ$ [$\boldsymbol{x}^\prime=(\boldsymbol{y}+\boldsymbol{x})/\sqrt{2}$], and $\varphi=-45^\circ$ [$\boldsymbol{y}^\prime=(\boldsymbol{y}-\boldsymbol{x})/\sqrt{2}$]. This anisotropy is a fingerprint of the gates layout. Indeed, there are larger areas not covered by metal gates in these two directions. As a consequence, the traps in these areas are more weakly screened and give rise to stronger electric fields that tend to squeeze the dots along either $\boldsymbol{x}^\prime$ or $\boldsymbol{y}^\prime$. The in-plane magnetic axes then align with the main axes of the resulting elliptical dot. As pointed out earlier, the principal $g$-factors of this elliptical dot are also different along the two in-plane axes. Notably, the variability is slightly larger along $\boldsymbol{x}^\prime$ and $\boldsymbol{y}^\prime$ than along $\boldsymbol{x}$ and $\boldsymbol{y}$ (see Figs. \ref{fig:g_factors_1}b and \ref{fig:g_factors_1}d), even though $\overline{g}^*$ is isotropic for in-plane magnetic fields. As shown in Ref. \cite{Mauro2024}, the effective $g$-factors are indeed more responsive when the magnetic field is aligned with the main electric field fluctuations.

\begin{figure}[t]
\centering
\includegraphics[width=0.98\columnwidth]{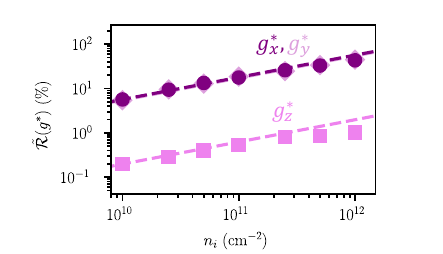}
\caption{Dependence of the relative IQR of the effective $g$-factors $g_x^*$, $g_y^*$ and $g_z^*$ on $n_i$. The dashed lines illustrate the $\tilde{\mathcal{R}}(f_R)\propto\sqrt{n_i}$ dependence expected from first-order perturbation theory.}
\label{fig:g_factors_3}
\end{figure}

We finally plot the relative IQR $\tilde{\mathcal{R}}(g^*)$ as a function of $n_i$ in Fig. \ref{fig:g_factors_3}, for $\boldsymbol{B}$ oriented along $\boldsymbol{x}$, $\boldsymbol{y}$ and $\boldsymbol{z}$. $\tilde{\mathcal{R}}(g^*)$ roughly scales as $\sqrt{n_i}$, as expected from perturbation theory \cite{Martinez2022}. $\tilde{\mathcal{R}}(g^*)$ shows by definition the same anisotropy as $\tilde{\mathcal{R}}(f_L)$. The relative variability of $g_z^*$ is thus more than one order of magnitude smaller than the relative variability of $g_x^*$ and $g_y^*$. Namely, $\tilde{\mathcal{R}}(g_z^*)$ remains below 1\% while $\tilde{\mathcal{R}}(g_x^*)$ and $\tilde{\mathcal{R}}(g_y^*)$ range between 4 and 30\% for trap densities from $n_i=10^{10}$ to $n_i=10^{12}$\,cm$^{-2}$. Note that a large spread of the effective $g$-factors may compromise the individual addressability of the qubits in large-scale architectures. We discuss the implications in Sec. \ref{sec:address}. 

\begin{figure}[t]
\centering
\includegraphics[width=0.98\columnwidth]{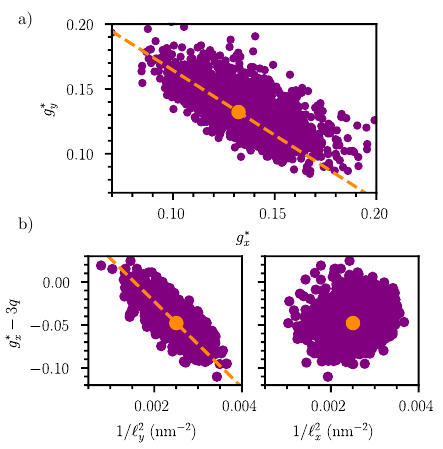}
\caption{a) Correlation between the two in-plane $g$-factors $g_x^*$ and $g_y^*$. The orange point is the data for the pristine device ($g_x^{*,0}=g_y^{*,0}=0.13$), and the dashed orange line highlights an anti-correlation with slope $-1$. b) Correlation between $g_x^*$ and $1/\ell_y^2$, $1/\ell_x^2$. The orange point is the data for the pristine device, and the dashed orange line is a guide to the eye with slope $-50$. The highlighted correlations show that the variability of $g^*$ essentially results from modulations of the QD size.}
\label{fig:g_factors_2}
\end{figure}

Interestingly, the variations of $g_x^*$ and $g_y^*$ anti-correlate (Fig. \ref{fig:g_factors_2}). We also observe clear correlations between $g_x^*$ and $1/\ell_y^2$ (and between $g_y^*$ and $1/\ell_x^2$). As a matter of fact,
\begin{align}
g_x^*&=\sqrt{g_{xx}^2+g_{yx}^2+g_{zx}^2} \nonumber \\
&=\sqrt{(3q+\delta g_{xx})^2+\delta g_{yx}^2+\delta g_{zx}^2} 
\end{align}
where $q=0.06$ is the cubic Zeeman parameter and $\delta g_{\alpha\beta}$  are corrections due to confinement and strains, given by Eq. (3) of Ref.~\cite{Mauro2024strain}. When these corrections are small, $g_x^*\approx 3q+\delta g_{xx}$ is dominated by the variations of $\delta g_{xx}$. In homogeneous strains, the latter are $\delta g_{xx}\approx-a\langle p_y^2\rangle+b\langle p_x^2\rangle\propto -a/\ell_y^2+b/\ell_x^2$ with $a\gg b$ (and $p_x$ and $p_y$ the momentum along $x$ and $y$). Likewise, $g_y^*\approx 3q+\delta g_{yy}$ where $\delta g_{yy}$ is $\propto - a/\ell_x^2+b/\ell_y^2$. We have also confirmed by calculations without cool-down strains (Appendix \ref{app:strains}) that the variability of the $g$-factors mainly results from the deformation of the dots rather than from their motion in the inhomogeneous strains, even though the latter are responsible for small rotations of the principal axes. As discussed in the next section, the inhomogeneous strains primarily give rise to a transverse coupling (Rabi oscillations) when the magnetic field goes in-plane.

All existing experiments on multiple Ge qubits show fingerprints of a significant variability of the effective $g$-factors. The four-qubit experiment of Ref. \cite{Hendrickx2021} reported $g^*=0.16$, $0.24$, $0.24$, $0.26$ for an in-plane magnetic field. More recently, a Relative Standard Deviation (RSD) of 70\% has been measured in a 10 QD device \cite{Wang2024}. While this is considerably larger than predicted in Fig. \ref{fig:g_factors_3}, in the experiment $\overline{g^*}=0.04$ is much smaller than in the simulations (which may be a consequence of smaller or anisotropic QDs). In fact, the standard deviation of $f_L$ is 392\,MHz/T in the experiment, which is compatible with the data reported above for $n_i>10^{11}$\,cm$^{-2}$. In a similar device and with a slightly off-plane magnetic field, $\overline{g^*}=0.58$ and the RSD drops to 5\% \cite{john2024}. This is again in line with the results shown here for $n_i\geq 10^{11}$\,cm$^{-2}$. Finally, the variability of the principal axes and values of $\hat{G}$ translates, at a given magnetic field orientation, in different precession axes for each qubit. This has also been observed in shuttling experiments \cite{vanRiggelen-Doelman2024}, and recently exploited to perform one-qubit shuttling gates \cite{Wang2024}. 

\subsubsection{Rabi frequencies}\label{sec:Rabi}

\begin{figure*}[t]
\centering
\includegraphics[width=0.98\textwidth]{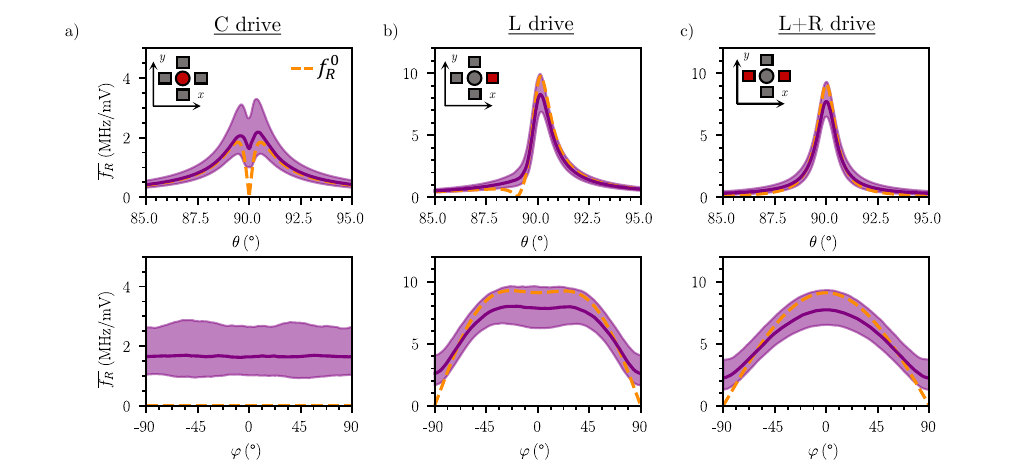}
\caption{Rabi frequency anisotropy in pristine and disordered devices. a) Median Rabi frequency as a function of $\theta$ (for $\varphi=0^\circ$) and $\varphi$ (for $\theta=90^\circ$) when driving with the C gate, for defective devices with trap density $n_i=10^{11}$\,cm$^{-2}$. The magnetic field amplitude is varied so that that the Larmor frequency is $f_L^0=2$\,GHz in the pristine device whatever the orientation. The dashed orange line is the Rabi frequency of the pristine device, and the shaded purple areas highlight the IQRs. b) Same for the L drive. c) Same for the L+R drive.}
\label{fig:fR_1}
\end{figure*}

In this section, we discuss the variability of the Rabi frequencies. Similar to previous studies \cite{Martinez2022-inhom,Abadillo2022}, we consider three kinds of resonant AC drives: with the plunger gate C ($V_{\rm C}=V_{\rm ac}\cos 2\pi f_L t$), with the left barrier gate L ($V_{\rm L}=V_{\rm ac}\cos 2\pi f_L t$), and with opposite modulations on the left and right barrier gates (L+R, $V_{\rm L}=-V_{\rm R}=\tfrac{1}{2}V_{\rm ac}\cos 2\pi f_L t$ ). While the two first are commonly used in experiments \cite{Hendrickx2021,Wang2024,john2024}, the L+R drive is interesting as it has the same symmetry as a homogeneous drive that displaces the QD as a whole (see Ref. \cite{Martinez2022-inhom} for details).

We compute the Rabi frequency $f_R$ from the derivative $\hat{g}^\prime$ of the $g$-matrix with respect to the drive \cite{Venitucci2018}:
\begin{equation}
    f_{R}=\frac{\mu_B V_\mathrm{ac}}{2h|\hat{g}\boldsymbol{B}|}\left|\hat{g}\boldsymbol{B}\times\hat{g}^\prime\boldsymbol{B}\right| \label{eq:fRg},
\end{equation}
where $V_\mathrm{ac}$ is the amplitude of the drive, which we set to $V_\mathrm{ac}=1$\,mV hereafter.

Recent studies have shown that the Rabi oscillations of hole spins in Ge can result from a wide variety of SOC mechanisms \cite{Martinez2022-inhom,Abadillo2022}. The cubic Rashba SOI \cite{Marcellina17,Terrazos21,Bosco2021} characteristic of quasi-2D dots is rather weak at small vertical electric fields. The Rabi oscillations are, therefore, usually dominated by the modulations of the principal $g$-factors (``conventional'' $g$-TMR) and principal axes of $\hat{G}$. The former result from the non-parabolic in-plane confinement and from the inhomogeneous vertical and drive fields, while the latter result from the coupling between the in-plane and out-of-plane motions of the hole \cite{Martinez2022-inhom}, and from the motion in the inhomogeneous shear strains imprinted by the differential thermal contraction of the materials \cite{Abadillo2022}. They are particularly efficient when the magnetic field is near the heterostructure plane. The effects of strains may supersede the other mechanisms in the absence of disorder \cite{Abadillo2022,Mauro2024strain}. The variability in position, shape and size of the QDs (Fig. \ref{fig:charge}), and in drive-induced displacements scatters the Rabi frequencies of the qubits. 

Fig. \ref{fig:fR_1} displays the median and IQR of the Rabi frequency as a function of the magnetic field angles $\theta$ (at $\varphi=0^\circ$) and $\varphi$ (at $\theta=90^\circ$) for the different AC drives. The dataset contains 2\,000 realizations of disorder with a charge trap density $n_i=10^{11}$\,cm$^{-2}$, and has been computed at constant $f_L^0=2$\,GHz (thus with an orientation-dependent magnetic field amplitude). We have added the pristine device data as the dashed orange lines for comparison.

\begin{figure}[t]
\centering
\includegraphics[width=0.98\columnwidth]{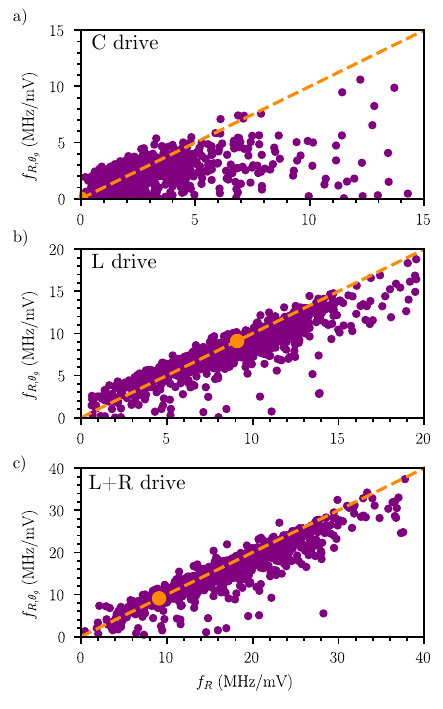}
\caption{Mechanisms of the Rabi oscillations. a-c) Correlations between the Rabi frequency $f_R$ of the disordered devices and the contribution $f_{R,\theta_g}$ from the modulations of $\theta_g$, for the C, L, and L+R drives. This contribution primarily results from the motion of the hole in the inhomogeneous shear strains. The magnetic field is oriented along $\theta=90^\circ$ and $\varphi=0^\circ$ ($\boldsymbol{B}\parallel\boldsymbol{x}$), and its amplitude is chosen so that $f_L^0=2$ GHz. The orange point is the pristine device data, and the dashed orange line is a guide-to-the-eye for $f_{R,\theta_g}=f_R$. The clear correlations show that the variability of $f_R$ is mainly dominated by inhomogeneous strains.}
\label{fig:fR_3}
\end{figure}

There are significant differences between the median and pristine Rabi frequencies. When driving with the C gate, the QD breathes but does not move in-plane in the absence of disorder. Owing to the symmetry of the device, the average inhomogeneous shear strains experienced by the hole remain zero. The only SOC mechanism at work is thus the modulation of the main $g$-factors arising from the variations of the dot size (conventional $g$-TMR). In a perfectly circular QD, the two principal in-plane $g$-factors remain degenerate, and consequently the Rabi frequency vanishes for in-plane magnetic fields \footnote{If the electric drive only gives rise to conventional $g$-TMR, $\hat{g}'=\mathrm{diag}(g_1',g_2',g_3')$. If $\boldsymbol{B}=(B_1,B_2,0)$, then $\|\boldsymbol{gB} \times \boldsymbol{g'B}\| = \left| g_1 g_2' B_1 B_2 - g_2 g_1' B_2 B_1 \right|$. In a circular QD $g_1=g_2$; and if it is perfectly centered under the C gate then $g_1'=g_2'$. Therefore, the in-plane Rabi frequency is zero. If the QD is not perfectly circular and/or not perfectly centered under the C gate, $f_R$ is only zero when $\boldsymbol{B}$ is parallel to the axes of $\boldsymbol{g_1}$ or $\boldsymbol{g_2}$.}. In the presence of disorder, however, the in-plane symmetry of the QD is broken, which enables conventional g-TMR for in-plane magnetic fields. Additionally, the dot is not necessarily centered under the gate, so that the C drive may displace the dot in the inhomogeneous strain field. The median Rabi frequency of the C gate is expected to be independent on $\varphi$. We find $\overline{f}_R/f_L^0\approx 1$\,MHz/mV/GHz at $\theta=90^\circ$. However, $\overline{f}_R$ rapidly decreases when $\theta$ goes out of plane due to the increase of $g^*$ (thus the decrease of the magnetic field amplitude at constant $f_L^0=2$\,GHz). The in-plane IQR is very large (see shaded area of Fig. \ref{fig:fR_1}a), as the disorder is responsible for the finite $f_R$. Variability significantly decreases when $\theta\gtrless 90^\circ$, which brings the optimal orientation slightly off-plane ($\theta \approx 87-89^\circ$ or $91-93^\circ$), where $\overline{f}_R$ is roughly the same but the IQR is smaller than at $\theta \approx 90^\circ$. This is in line with the conclusions of Ref. \cite{john2024}, where the authors also identified off-plane magnetic fields as the optimal operation point. Note that Ref. \cite{john2024} also showed that the C gate is more efficient in three-hole qubits. 

For L and L+R drives, the angular dependence of the Rabi frequency is also smoothed out in the presence of traps, especially near the zeros of the Rabi frequency, which result from symmetry considerations (broken by the disorder) or from perfect cancellations between different mechanisms (also blurred by disorder). The median Rabi frequencies are slightly smaller than in the pristine device and reach $\overline{f}_R/f_L^0\approx 5$\,MHz/mV/GHz for $\boldsymbol{B}\parallel\boldsymbol{x}$ ($\theta=90^\circ$, $\varphi=0^\circ$). The IQR is also maximal in plane, and is weakly dependent of $\varphi$ ($\mathcal{R}(f_R)/\overline{f}_L \approx 1.7$\,MHz/mV/GHz at $n_i=10^{11}$\,cm$^{-2}$). The optimal magnetic field orientation for the L and L+R drives is $\theta\to 90^\circ$ and $\varphi\to 0^\circ$ where $f_R$ is maximal.  

In order to disclose the dominant driving mechanism, we split the total $f_R$ into different contributions. We first diagonalize the $g$-matrix $\hat{g}={\rm diag}(g_1, g_2, g_3)$ of each qubit by choosing an appropriate spin basis set and the principal axes for the magnetic field coordinates \cite{Venitucci2018}. We next split the $\hat{g}'$-matrix in this representation into three blocks:
\begin{subequations}
    \begin{align}
    \hat{g}'_{\rm TMR} &=\begin{bmatrix} g'_{11} & 0  & 0 \\ 0 & g'_{22}  & 0  \\  0 & 0  & g'_{33} \end{bmatrix}\,,  \\
    \hat{g}'_{\theta_g} &=\begin{bmatrix} 0 & 0  & g'_{13} \\ 0 & 0  & g'_{23}  \\  g'_{31} & g'_{32}  & 0 \end{bmatrix}\,,  \\
    \hat{g}'_{\varphi_g} &=\begin{bmatrix} 0 & g'_{12}  & 0 \\ g'_{21} & 0  & 0  \\  0 & 0  & 0 \end{bmatrix}\,.
\end{align}
\end{subequations}
The first one, $\hat{g}'_{\rm TMR}$, is the conventional $g$-TMR  resulting from the electrical modulations of the principal $g$-factors. The second one, $\hat{g}'_{\theta_g}$, collects the off-diagonal elements $g'_{13}$, $g'_{31}$, $g'_{23}$ and $g'_{32}$ that primarily capture the variations of the tilt $\theta_g$ of the out-of-plane principal axis (and the Rashba SOI) \cite{Abadillo2022,Martinez2022-inhom}. The third one, $\hat{g}'_{\varphi_g}$, essentially accounts for the variations of the orientation $\varphi_g$ of the in-plane principal axes. The total $\hat{g}'$ matrix is therefore $\hat{g}'=\hat{g}'_{\rm TMR}+\hat{g}'_{\theta_g}+\hat{g}'_{\varphi_g}$. We can thus compute $\boldsymbol{f}_{R,\alpha}$ and $f_{R,\alpha}=|\boldsymbol{f}_{R,\alpha}|$ using $\hat{g}'_\alpha$ in Eq.~\eqref{eq:fRg} for $\alpha\in[\rm TMR,\,\theta_g,\,\varphi_g]$. Note that the individual contributions can make constructive or destructive interferences as $f_R=|\boldsymbol{f}_{R,\rm TMR}+\boldsymbol{f}_{R,\theta_g}+\boldsymbol{f}_{R,\varphi_g}|$.

We plot in Fig. \ref{fig:fR_3}a the correlation between $f_R$ and $f_{R,\theta_g}$ for the C drive and $\boldsymbol{B}\parallel\boldsymbol{x}$. The modulations of $\theta_g$ resulting from the motion in the inhomogeneous shear strains make the main contribution to $f_R$ for in-plane magnetic fields. Yet $f_R<f_{R,\theta_g}$ for several devices because the interference between the shear strains and the other mechanisms is usually destructive for the C drive.

For L and L+R drives, the Rabi oscillations are clearly ruled by $\hat{g}'_{\theta_g}$, and, more specifically, by the large $g'_{13}$ resulting from the motion of the dot in the inhomogeneous shear strains (see Figs. \ref{fig:fR_3}b and \ref{fig:fR_3}c, where $f_{R,\theta_g}$ and $f_R$ show a remarkable correlation with slope 1). ${\overline f}_R$ is maximal at $\theta=90^\circ$, and shows an approximate $\rm cos(\varphi)$ dependence for in-plane magnetic fields, another fingerprint of the dominant $g_{13}'$. Therefore, the electrical modulation of $\theta_g$ by strain is still the prevalent driving mechanism when the QD is shaken in a disordered potential. Moreover, $\boldsymbol{f}_{R,\theta_g}$ is mostly orthogonal to $\boldsymbol{f}_{R,\rm TMR}$ and $\boldsymbol{f}_{R,\varphi_g}$, which tend to be antiparallel and cancel each other. 

\begin{figure}[t]
\centering
\includegraphics[width=0.98\columnwidth]{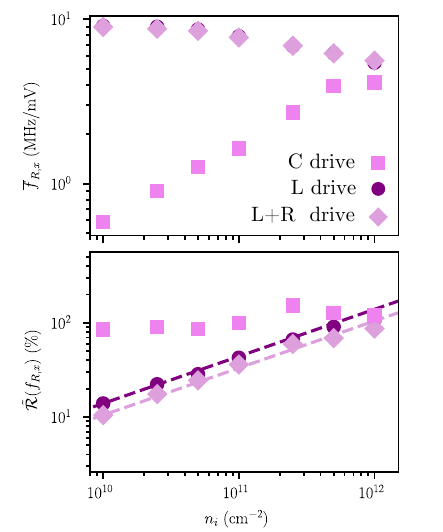}
\caption{a) Median Rabi frequency $\overline{f}_R$ and b) relative IQR $\tilde{\mathcal{R}}(f_R)$ as a function of $n_i$ when driving with the C gate (squares), L gate (circles) and L+R gates (diamonds). The magnetic field is oriented along $\theta=90^\circ$ and $\varphi=0^\circ$ ($\boldsymbol{B}\parallel\boldsymbol{x}$), and its amplitude is chosen so that $f_L^0=2$ GHz. The dashed lines highlight the $\tilde{\mathcal{R}}(f_R)\propto\sqrt{n_i}$ dependence expected from first-order perturbation theory. }
\label{fig:fR_2}
\end{figure}

We plot in Fig. \ref{fig:fR_2} the median ${\overline f}_{R,x}$ and the relative IQR $\tilde{\mathcal{R}}(f_{R,x})$ of the Rabi frequency for $\boldsymbol{B}\parallel\boldsymbol{x}$ as a function of the trap density. The variability of the C drive is the largest (around 100\% whatever $n_i$). This results from the fact that the Rabi oscillations are promoted by disorder (${\overline f}_{R,x}$ and $\sigma (f_{R,x})\rightarrow 0$ when $n_i\rightarrow 0$). The L and L+R drives show similar variabilities, ranging from 90\% for $n_i=10^{12}$\,cm$^{-2}$ down to 10\% for $n_i=10^{10}$\,cm$^{-2}$. At variance with the C drive, ${\overline f}_{R,x}$ is weakly dependent on $n_i$ and $\tilde{\mathcal{R}}(f_{R,x})$ follows the $\propto\sqrt{n_i}$ dependence expected from first-order perturbation theory \cite{Martinez2022}. These results emphasize that the side gates achieve both larger Rabi frequencies and lower variability (for single holes). Surprisingly, these levels of variability are not much smaller than reported for Si MOS despite the lower disorder in the vicinity of the qubits \cite{Martinez2022}. It is nonetheless worth noting that inhomogeneous strains are not accounted for in Ref. \cite{Martinez2022}, but appear to be the main driving mechanisms here. 

The Rabi frequencies reported above are consistent with previous simulation studies using similar geometries \cite{Abadillo2022,Mauro2024} and with the existing experimental data. In Ref. \cite{Hendrickx2021}, $\overline{f}_R/\overline{f}_L=5.5$\,MHz/mV/GHz for in-plane magnetic fields and drives equivalent to L. In the 10 qubits experiment of Ref. \cite{john2024}, the authors systematically characterized $f_R$ for the C and L drives. For $\theta\approx 87^\circ$ and $\varphi\approx 45^\circ$, they measure $\overline{f}_R/\overline{f}_L=0.14$\,MHz/mV/GHz and 0.18 MHz/mV/GHz for the C and L drive, respectively. Additionally, the RSD reaches roughly 50\% for both drives. These results are qualitatively compatible with those reported here for $n_i\geq 10^{11}$\,cm$^{-2}$ \footnote{We obtained, for $n_i = 10^{11}$\,cm$^{-2}$ and $\theta=87^\circ$, $\phi=0^\circ$, $\overline{f_R}/\overline{f_L} = 0.39$ MHz/mV/GHz for the C drive, and $\overline{f_R}/\overline{f_L} = 0.42$ MHz/mV/GHz for the C drive. The RSD are 65\% and 58\% for the C and L drive, respectively.}.

\subsubsection{Dephasing}\label{sec:sweet}

We next study the anisotropy of charge-noise-driven dephasing in the presence of static disorder. We emphasize that the defects responsible for charge noise and for variability are in principle different. On the one hand, variability is due to deep traps at the SiGe/oxide interface that rarely undergo trapping/detrapping events. On the other hand, dephasing is due to shallower traps whose state switches during the experiment (e.g., through charge transfers between the traps and the gates above or charge hopping between neighboring traps). The electric field generated by these two-level fluctuators (TLFs) modulates the vertical and in-plane confinement of the quantum dot, as do the gates. In a first approximation, we can therefore lump the effects of the TLFs into effective gate voltage modulations (provided the gates sample a relevant set of electrical perturbations), and estimate the dephasing rate as \cite{Mauro2024,bassi2024}
\begin{equation}\label{eq:Gamma2}
    \Gamma^*_2 = \frac{1}{T^*_2} = \sqrt{2} \pi \sqrt{\sum_{\alpha} \left[ \delta V^{\rm rms}_{\alpha} \frac{\partial f_L}{\partial V_{\alpha}} \right]^2}\,,
\end{equation}
where $\alpha\in [\textrm{C,\,L,\,R,\,B,\,T}]$ \footnote{We have restricted the contributions to the C and side gates, as the contribution from the second nearest neighbors is found to be negligible.}, $V^{\rm rms}_{\alpha}$ is the rms amplitude of the noise on gate $\alpha$, and ${\partial f_L}/{\partial V_{\alpha}}$ is the longitudinal spin electric susceptibility (LSES) of gate $\alpha$ \cite{Piot2022,bassi2024}. In this study, we set $\delta V^{\rm rms}_{\alpha}=10$\,$\mu$V for all gates, which yields values of $\Gamma_2^*$ that are consistent with experiments. Note that this simple model relies on quite strong assumptions, as it can only capture the effect of TLFs that are close to the gates, but may fail to describe the effect of TLFs located between the gates. The general trends discussed below are not, however, expected to be dependent on the details of the model. We discuss the limitations of this model in more details hereafter.

\begin{figure}[t]
\centering
\includegraphics[width=0.98\columnwidth]{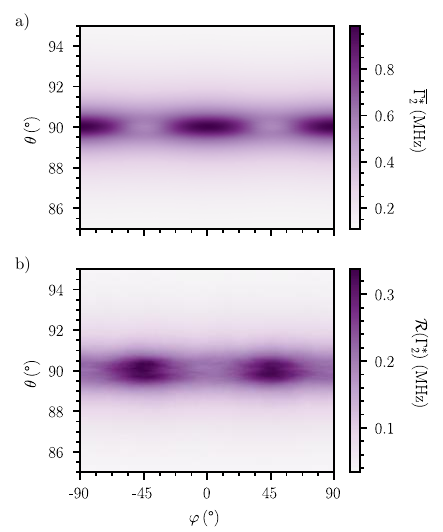}
\caption{Dephasing in the presence of disorder. (a) Median and (b) IQR of the dephasing rate $\Gamma_2^*$ as a function of the magnetic field orientation for $n_i=10^{11}$\,cm$^{-2}$. The amplitude of the magnetic field is chosen so that $f_L^0=2$ GHz.}
\label{fig:sweet_maps2}
\end{figure}

We compute the LSES and estimate $\Gamma_2^*$ with Eq.~\eqref{eq:Gamma2} for each defective device, and plot the median dephasing rate $\overline{\Gamma}_2^*$ and IQR $\mathcal{R}({\Gamma}_2^*)$ over 2000 samples of disorder in Fig. \ref{fig:sweet_maps2}. 

In the absence of disorder, $\Gamma^*_2$ is minimal for $\theta=0^\circ$ and maximal for $\theta=90^\circ$ as the in-plane $g$-factors are more sensitive to noise \cite{Mauro2024}. $\Gamma^*_2$ exhibits, nevertheless, local optimums for in-plane magnetic fields at $\varphi=\pm 45^\circ$. The median dephasing rate $\overline{\Gamma}_2^*$ of the disordered devices shows the same anisotropy. As discussed in Section \ref{sec:Rabi}, the Rabi frequency is sizable only when $\theta\approx 90^\circ$, so that the domains $\theta\to 0^\circ$ and $\theta\to 180^\circ$ where $\overline{\Gamma}_2^*$ decreases drastically are hardly exploitable experimentally. We discuss this matter in more details in Section \ref{sec:Qfact}. With $V^{\rm rms}_{\alpha}=10$\,$\mu$eV, $\overline{\Gamma}_2^*$ reaches $\approx 1$\,MHz for $\theta=90^\circ$ and $\varphi=0^\circ$. We emphasize that the two in-plane minima at $\varphi=\pm 45^\circ$ are highly dependent on the TLFs model. They result in the present description from the fact that the gates hardly capture the effects of electric field fluctuations along the diagonals of the device (the $[100]$ and $[010]$ axes). The electric field noise shall actually be stronger along $x=[1\bar{1}0]$ and $y=[110]$ if the traps exchange charges with the gates (as only traps below the gates can then switch). This assumption breaks down, however, if hopping or reconfiguration of traps between the gates play a major role (especially because they are not screened by the gates). The contrast between the in-plane extrema of $\overline{\Gamma}_2^*$ along $[100]/[010]$ and along $[1\bar{1}0]/[110]$ is thus highly dependent on the nature and distribution of the traps (largely unknown at this stage), but the overall dependence on $\theta$ is expected to be solid.

The IQR of $\Gamma_2^*$ is also maximal in-plane at $\varphi=\pm 45^\circ$, and minimal at $\varphi=0^\circ$. This dependence is, again, a fingerprint of the gate layout and model for charge noise.

\begin{figure}[t]
\centering
\includegraphics[width=0.98\columnwidth]{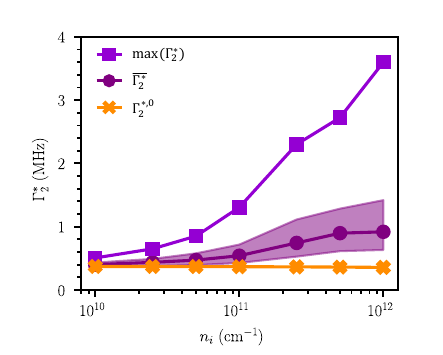}
\caption{Dependence of $\Gamma_2^*$ on $n_i$. $\overline{\Gamma}_2^*$ is plotted with purple dots, with the IQR highlighted by the shaded area. The pristine device value ($\Gamma_2^{*,0}$) is plotted with orange crosses, and the maximum $\Gamma_2^*$ among the 95\% best devices with violet squares. The magnetic field is oriented along $\theta=90^\circ$, $\varphi=45^\circ$, and its amplitude is chosen so that $f_L^0=2$ GHz.}
\label{fig:sweet_dens}
\end{figure}

In Fig. \ref{fig:sweet_dens} we plot $\overline{\Gamma}_2^*$ as a function of $n_i$ at $\theta=90^\circ$, $\varphi=45^\circ$, as well as the IQR (shaded purple area), the value in the pristine device, and the maximal $\Gamma_2^*$ among the 95\% best devices. $\overline{\Gamma}_2^*$ shows a sizable increase with $n_i$, which is not observed for the pristine device. The IQR of $\Gamma_2^*$ also increases significantly, and reaches $\mathcal{R}(\Gamma_2^*)=0.9$\,MHz at $n_i=10^{12}$\,cm$^{-2}$. These two observations are in fact tightly related: as $\Gamma_2^*$ is positive by definition, but the LSES in Eq.~\eqref{eq:Gamma2} are signed, a large variability of the latter yields to an increase of both the median and IQR of the former. Finally, $\rm max(\Gamma_2^*)$ also illustrates the tailed nature of the distribution of $\Gamma_2^*$. As defined here, $\rm max(\Gamma_2^*)$ is nothing else than the 95th percentile, and is significantly larger than the value $1.22\mathcal{R}(\Gamma_2^*)$ expected for normal distributions. It is a relevant metric because the performance of a large scale quantum processor are limited by the qubits with fastest decoherence. We explore the constraints that this implies in Section \ref{sec:Qfact}. 

In the appendices, we provide additional data on the dependence of the variability on $V_{\rm C}$ (the confinement strength, Appendix \ref{app:depVc}), and on the top GeSi barrier thickness $H_{\rm SiGe}$ (Appendix \ref{app:thick}). We show, as expected, that increasing $H_{\rm SiGe}$ (thus bringing the traps farther away from the Ge well) reduces variability.

\section{Implications of variability for arrays of spin qubits} \label{sec:implications}

In this Section we discuss the implications of the results discussed above on the performance of large-scale spin qubit architectures. In small spin qubit arrays, the variability can be managed by individually retuning the qubits. In particular, with robust lever-arms, adjustments of the gate voltages can compensate for the variability in $\mu$ shown in Fig. \ref{fig:charge}. The feasibility of such a procedure has been extensively discussed in Ref. \cite{Martinez2024}. It is unclear, however, whether the large dispersion of $g^*$ reported in Fig. \ref{fig:g_factors_1} can be balanced by gate voltage corrections. The variability of the Rabi frequency can also be corrected by individually adjusting the drive amplitude $V_{\rm ac}$, as done for example in Ref. \cite{Philips2022}. Large-scale architectures, however, may set constraints and limitations on individual tuning. In particular, crossbar architectures without individual plunger gates for each qubit must be robust against the predicted $\Delta \mu$'s. Moreover, the management of individual $V_{\rm ac}$'s can quickly become challenging.

In Subsection \ref{sec:crossbar} we discuss how the variability in $\mu$ can compromise the functionality of crossbar architectures. In Subsection \ref{sec:address}, we address the constraints on single-qubit addressability imposed by the variability in $g^*$. Finally, in Subsection \ref{sec:Qfact} we explore how the variability in $f_R$ and $\Gamma_2^*$ can compromise the performance of a large-scale quantum processor. 

\subsection{Crossbar architectures} \label{sec:crossbar}

Spin qubit platforms will inevitably face routing challenges as they scale up. These challenges are more difficult to tackle when the pitch is small and the number of control knobs per qubit is large. One proposed strategy to address this problem is to implement shared control schemes, such as the so-called crossbar architectures \cite{Veldhorst2017,Vinet2018,Vandersypen2017}. A $4\times 4$ crossbar array has actually been fabricated and characterized in Ge/SiGe heterostructures, and odd charge occupancy achieved in all QDs \cite{Borsoi2023}.  

In crossbar architectures, several qubits generally share the same gate potential. It is clear, therefore, that disorder-induced variations of the chemical potential $\mu$ can impede the operation of the devices. In particular, single-hole occupancy can be compromised in the qubits whose $|\Delta\mu|$ is larger than $U/2$. As a matter of fact, not all QDs could be biased in the single-hole regime in the crossbar array of Ref. \cite{Borsoi2023}. A key condition for reliable crossbar operation is thus $|\Delta\mu|<U/2$ for all qubits. 

\begin{figure}[t]
\centering
\includegraphics[width=0.98\columnwidth]{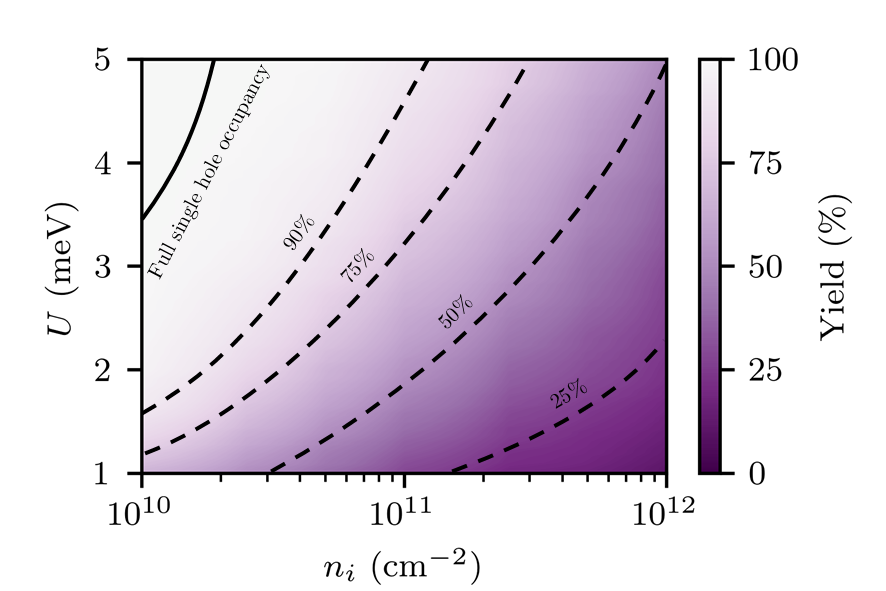}
\caption{Fraction of devices fulfilling $|\Delta\mu|<U/2$ as a function of the charging energy $U$ and trap density $n_i$.}
\label{fig:crossbar}
\end{figure}

$\mathcal{R}(\mu)$ and $\mathcal{R}(U)$ are plotted as a function of $n_i$ in Fig. \ref{fig:charge}. As discussed before, $\mathcal{R}(U)$ is much smaller than $\mathcal{R}(\mu)$. Therefore, we consider hereafter $U$ as a deterministic parameter that defines the margins of operation. We monitor the fraction of devices that fulfill $|\Delta\mu|<U/2$ in the datasets of Fig. \ref{fig:charge}b, and plot this yield as a function of $n_i$ and $U$ in Fig. \ref{fig:crossbar} \footnote{We emphasize that $U$ may also be tuned with $V_{\rm C}$, and that $\mathcal{R}(\mu)$ shows only a weak dependence on gate voltage (see Appendix \ref{app:depVc}).}.

Single hole occupancy can be achieved in $>90\%$ of the devices for large enough $U$ and low enough $n_i$. However, the median charging energy of the present QDs is only $\overline{U}\approx2.4$\,meV. Reaching such yields for $U<2.5$\,meV actually demands very good interface qualities $n_i\lesssim 3\times 10^{10}$\,cm$^{-2}$. We achieve full single hole occupancy in the sets of 1\,000 to 2\,000 simulated devices only for $U>3.5$\,meV when $n_i\to 10^{10}$\,cm$^{-2}$, which is at the state-of-the-art of micro-electronics technologies (yet for Si/SiO$_2$, not for SiGe/Al$_2$O$_3$ interfaces) \cite{Bauza02,Brunet09,Pirro16,Vermeer21}. We emphasize, though, that the ``full single hole occupancy'' line on Fig. \ref{fig:crossbar} is indicative and not statistically meaningful (as outliers can still appear in different or larger samples of devices). These results show that crossbar architectures may indeed suffer from the variability in $\mu$, and that reliable functionality would require large $U$ and/or a substantial improvement in interface quality. Alternatively, specific gate layouts have been proposed to cope with the variability in $\mu$, and could ease the implementation of such architectures \cite{Martinez2024}.

\subsection{Individual qubit addressability} \label{sec:address}

Another challenge for scalability is qubit addressability. In arrays of spin qubits, one-qubit gates may rely on the dependence of $f_L$ on the electrical confinement (the Stark shift) to bring the qubits on and off resonance with a single (or a few) RF generator(s) with fixed frequency. This protocol is certainly feasible if all qubits have the same $f_L$ and LSES, yet a large spread of these quantities may hinder its implementation. Note that due to restrictions in space and cooling power, including many RF generators on the chip is hardly possible. 

In the present devices the electrical confinement is primarily controlled by the difference of potential $V_{\rm C}$ between the central and side gates. In order to assess the addressability of the qubits, we assume a single drive with frequency $f_0=\overline{f_L}(V_{\rm C}^0)$, and define a bias window $\delta V_{\rm C}=V_{\rm C}-V_{\rm C}^0\in[\delta V_{\rm C}^{\rm min},\,\delta V_{\rm C}^{\rm max}]$ for device tuning (with $V_{\rm C}^0$ the original bias such that $r^0_\parallel=28.3$\,nm). We then count the fraction (yield) of qubits whose Larmor frequency $f_L$ can be matched to $f_0$ within this window. For that purpose, we compute the $g$-matrix $\hat{g}$ of each qubit at four different $\delta V_{\rm C}=-40$, $-20$, $0$, $20$\,mV and fit $f_L(\delta V_{\rm C})$ to a polynomial for each magnetic field orientation. 

\begin{figure}[t]
\centering
\includegraphics[width=0.98\columnwidth]{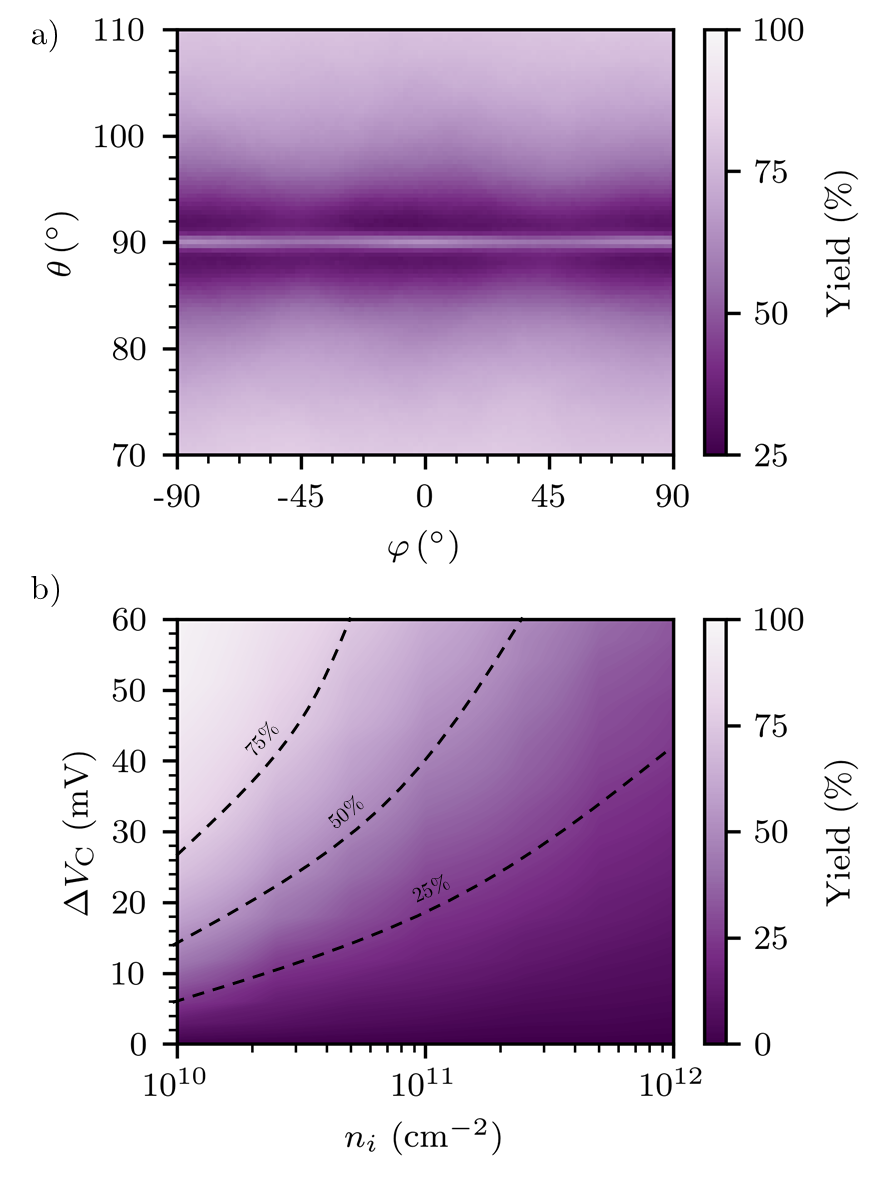}
\caption{a) Fraction of the devices whose Larmor frequency can be tuned to a target $f_L=\overline{f_L}$, as a function of the magnetic field orientation, allowing for a $\Delta V_{\rm C}=60$\,mV bias corrections window. The trap density is $n_i=10^{11}$\,cm$^{-2}$. b) Fraction of the devices that can be tuned to the target $f_L=\overline{f_L}$ as a function of the bias window $\Delta V_{\rm C}$ and $n_i$, for a magnetic field $\boldsymbol{B}\parallel\boldsymbol{x}$.}
\label{fig:addr_map}
\end{figure}

Note that constraints apply to $\delta V_{\rm C}$, as the central gate potential must remain attractive enough to form the QD, yet shall not be too attractive to prevent the hole from escaping to the GeSi/Al$_2$O$_3$ interface. Additionally, large negative $\delta V_{\rm C}$ require a tight control over the barriers between the qubits to hold the quantum dot occupancy during manipulation. However, a given negative $\delta V_{\rm C}$ is roughly equivalent to a positive $\delta V=-\delta V_{\rm C}$ on all side gates; it is, therefore, in principle possible to tune the confinement with minimal chemical potential variations across the dots using appropriate virtual gates (except in crossbar architectures). Taking these constraints into account, we consider asymmetric bias windows $\delta V_{\rm C}\in[-2/3\Delta V_{\rm C},1/3\Delta V_{\rm C}]$. The map of the yield is plotted as a function of the orientation of the magnetic field in Fig. \ref{fig:addr_map}a for $n_i=10^{11}$\,cm$^{-2}$ and $\Delta V_{\rm C}=60$\,mV. With lever-arms of the order of 0.2 eV/V, $\Delta V_{\rm C}=60$\,mV is already $>4$ times larger than $U$.

The yield is maximum (64\%) at $\theta=90^\circ$, where, surprisingly, the variability of $g^*$ also peaks (see Sec. \ref{sec:gfact}). Indeed, the yield decreases with ${\cal R}(f_L)$, but increases with the LSES of the C gate (better tunability). In the present devices, the decrease of the median LSES of gate C (at constant Larmor frequency) is faster than the decrease of $\tilde{\cal R}(f_L)$ when the magnetic field goes slightly out-of-plane \cite{Mauro2024}. For $\theta=90\pm 2^\circ$, the fraction of tunable devices drops to 30\%, which highlights the importance of a tight control over the orientation of $\boldsymbol{B}$. At large $\theta$, $\tilde{\cal R}(f_L)$ however decreases faster than the LSES, and the yield increases again. As discussed previously, this region is hardly exploitable due to small Rabi frequencies. For in-plane magnetic fields, the yield is slightly smaller (54\%) at $\varphi=\pm 45^\circ$ because the  IQR of the $g$-factors are slightly larger than at $\varphi=0^\circ$. The optimal magnetic field orientation for best qubit addressability is, therefore, $\boldsymbol{B}\parallel\boldsymbol{x}$.

We next set the magnetic field along this optimal orientation and plot the yield as a function of $n_i$ and $\Delta V_{\rm C}$ in Fig. \ref{fig:addr_map}b. Good addressability requires large bias corrections even for moderate interface disorder. There are still $\sim$3\% of the devices that cannot be tuned at $n_i=10^{10}$\,cm$^{-2}$ and $\Delta V_{\rm C}=60$\,mV. The situation worsens for larger $n_i$, as the yield is 64\% for $n_i=10^{11}$\,cm$^{-2}$, and 33\% for $n_i=10^{12}$\,cm$^{-2}$. As discussed above, $\Delta V_{\rm C}=60$\,mV is already significantly larger than $U$, yet for smaller bias windows and standard interface qualities ($n_i=10^{11}-10^{12}$\,cm$^{-2}$), the yield barely reaches 50\%.  

In view of these results, it seems quite inevitable that each Ge qubit has to a large extent its own ``spin personality''. In this respect, protocols trying to tune the qubits to a fixed RF signal may be challenging even at low $n_i$. Tuning the RF to each qubit may, on the other hand, compromise parallel operation. Spin manipulation by shuttling \cite{Wang2024} appears, therefore, as a promising approach that turns the differences between the qubits into a resource, at the possible price of larger variability in the Rabi frequencies. In any case, it is clearly essential to take into account the variability of $f_L$ when designing protocols for spin manipulation and shuttling in large-scale Ge architectures.

\subsection{Processor quality factors} \label{sec:Qfact}

\begin{figure*}[t]
\centering
\includegraphics[width=0.98\textwidth]{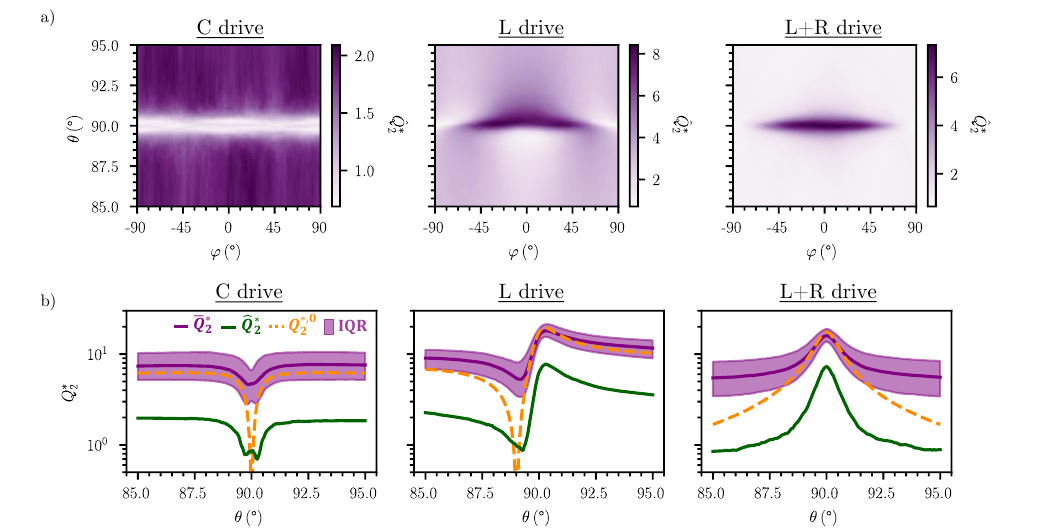}
\caption{Quality factors in disordered devices. a) Quality factor $\hat{Q}_2^*$ of the quantum processor as a function of the magnetic field orientation for $n_i=10^{11}$\,cm$^{-2}$ and a drive with the C, L and L+R gates. b) Median $\overline{Q}_2^*$ and IQR $\mathcal{R}({Q}_2^*)$ of the quality factor of the qubits, quality factor $\hat{Q}_2^*$ of the processor, and quality factor $Q_2^{*,0}$ of the pristine device as a function of the angle $\theta$ of the magnetic field ($\varphi=0^\circ$), for $n_i=10^{11}$\,cm$^{-2}$ and the same three drives as in a).}
\label{fig:sweet_maps}
\end{figure*}

In this Subsection, we analyze the impact of the variability of $\Gamma_2^*$ and $f_R$ onto the performance of a large-scale quantum processor. The coherence of such a processor would be limited by the qubits with the largest dephasing rates $\Gamma_2^*$, whereas the operation times would be limited by the qubits with the slowest Rabi frequencies $f_R$ (assuming that the two-qubit gates are fast enough not to limit the performances). Note that the slowest qubits and the ones with fastest decoherence are generally not the same. We can define the quality factor of a qubit as the number of $\pi$ rotations that can be achieved within the dephasing time,
\begin{equation}
    Q^*_2 = 2 f_R/\Gamma_2^*.
\end{equation}
It is thus clear that the variability of $f_R$ and $\Gamma_2^*$ translates into a variability of $Q^*_2$. Along the lines of Ref. \cite{Martinez2022}, we also define a quality factor for the whole quantum processor. For that purpose, we assume that we can discard the worst-performing qubits (the $5\%$ slowest, and the $5\%$ with fastest dephasing). With the remaining qubits, we compute
\begin{equation}
    \hat{Q}^*_2 = \textrm{min}(f_R)/\textrm{max}(\Gamma_2^*),
\end{equation}
which is always smaller than $\overline{Q}_2^*$.

In Fig. \ref{fig:sweet_maps}a, we plot $\hat{Q}_2^*$ as a function of the magnetic field orientation for the C, L and L+R drives and a charge trap density $n_i=10^{11}$\,cm$^{-2}$. In Fig. \ref{fig:sweet_maps}b, we also plot $\hat{Q}_2^*$, the median and IQR of $Q_2^*$, and the quality factor $\hat{Q}_2^{*,0}$ of the pristine device as a function of the angle $\theta$ for $\varphi=0^\circ$. The C drive shows the smallest $\overline{Q}_2^*$ owing to the slow $f_R$, and exhibits a significant IQR, fingerprint of the large variability in $f_R$ (see Fig. \ref{fig:fR_1}). Moreover, the optimal operation point for this drive is in fact slightly out of plane ($\hat{Q}_2^*=2$ for $\theta=95^\circ$ and $\varphi=45^\circ$, as $\hat{Q}_2^*$ drastically drops down to $\approx 1$ for $\theta\to 90^\circ$). This is the consequence of having at least one qubit with low enough disorder so that $f_R\approx 0$ for in-plane magnetic fields. In the absence of disorder, the in-plane quality factor is indeed zero as $f_R\to0$.

The L and L+R drives perform similarly. The anisotropy of $Q_2^*$ is dominated by the variations of $f_R$, which peaks in-plane. $\overline{Q}_2^*$ shows the same qualitative behavior as the pristine device, and reaches 17 for the L and 16 for the L+R drive at $\theta=90^\circ$, $\varphi=0^\circ$. The IQR also highlights a significant spread of the quality factors, with $\mathcal{R}({Q}_2^*) \approx 5.5$ for both drives at the same magnetic field orientation. The maximal $\hat{Q}_2^*$ is achieved at $\theta=90^\circ$ and is 7.7 for L and 7.2 for L+R drives. The optimal magnetic field orientation for these drives is therefore in-plane, which ensures both best qubits on average and the best performance for the quantum processor as a whole. 

To conclude this section, we would like to outline several efficient strategies to mitigate variability. First, the results discussed in this work unambiguously show that improving materials (decreasing $n_i$) is the most effective way to reduce variability. Yet achieving qubit homogeneity (within a few percents) would require interface qualities much beyond the state-of-the-art. Besides this option, we show in Appendix \ref{app:thick} that thick top SiGe barriers also help to improve the homogeneity of the qubits while still enabling a decent electrical control. Finally, the engineering of gate layouts \cite{Martinez2022} and strains (see Appendix \ref{app:strains}) also provide solutions to reduce variability. Nevertheless, while reducing variability will certainly help scaling spin qubit technologies, quantum architectures must be able to deal with an inevitable level of device inhomogeneity.

\section{Conclusions} 

In this study, we have quantified the variability of the charge and spin properties of Ge spin qubits resulting from interface traps at the top SiGe/oxide interface. Our analysis reveals a moderate variability of charge properties, with ensemble values notably better than those reported for Si MOS devices. However, the variability of spin properties remains significant. This is primarily attributed to the diverse spin-orbit coupling mechanisms at work in Ge and, in particular, to the pronounced sensitivity of in-plane $g$-factors to modulations of the QD size and shape, and of $f_R$ to local strains. We find that driving the qubits with one (or two) side gates is more robust against disorder than driving with the plunger gate; yet achieving variabilities below 10\% in both the Larmor and Rabi frequencies calls for trap densities $n_i\lesssim 10^{10}$\,cm$^{-2}$.

We have also assessed the implications of variability. We have explored the effect of chemical potential fluctuations on the operation of crossbar architectures. Despite the relatively low variability of this chemical potential, our findings indicate that crossbar architectures may require state-of-the-art interface quality to ensure single-hole occupancy in all qubits. The variability of in-plane $g$-factors also poses a significant challenge for manipulation protocols relying on Stark shifts to bring qubits in and out of resonance with a RF signal. This highlights the need to account for this variability in the design of protocols for large-scale spin manipulation. Similarly, the distribution of Rabi frequencies and decoherence times can impact the performance of quantum processors. We find that the optimal magnetic field orientation for Ge qubits, that minimizes the impact of disorder and maximizes the quality of the qubits and quantum processor is in plane when driving with side gate(s), and a few degrees out-of-plane when driving with the plunger gate.

Overall, this study underscores the critical importance of material and interface quality in achieving homogeneous and reproducible qubits. Despite the lower intrinsic disorder in Ge/GeSi heterostructures, variability remains a significant challenge when scaling up quantum devices. The vision of large-scale quantum architectures with homogeneous qubits appears, therefore, little realistic. Instead, each qubit shows its own personality to some extent, and quantum manipulation protocols and hardware must be able to cope with this variety, and even turn it into a resource \cite{Wang2024}.

\section*{Acknowledgements}

We thank Cécile Yu, Esteban A. Rodriguez-Mena, and José-Carlos Garcia Abadillo-Uriel for fruitful discussions. This work was supported by the ``France 2030'' program (PEPR PRESQUILE-ANR-22-PETQ-0002), the French National Research Agency (project InGeQT), and by the Horizon Europe Framework Program (grant agreement 101174557 QLSI2).

\appendix

\section{Surface roughness variability}\label{app:SR}

In this work, we have focused on charge traps at the SiGe/oxide interface as the source of disorder. Indeed, epitaxial interfaces are expected to be free of such traps, but they may be interdiffused and rough. The correlation length ($L_\mathrm{C}\sim 50-100$\,nm) and rms amplitude ($\delta z_\mathrm{rms}\sim 0.2-0.7$\,nm) of this roughness have been characterized for Si/SiGe heterostructures \cite{peña2023}. $L_\mathrm{C}$ is thus comparable or larger than the typical dot size ($2r_\parallel\approx 56$\,nm in this study), so that the interface can be considered, in a first approximation, almost flat at the scale of each QD. Nonetheless, such roughness gives rise to qubit-to-qubit fluctuations of the quantum well thickness at larger scale, and may thus contribute to the variability in arrays. Although $L_\mathrm{C}$ and $\delta z_\mathrm{rms}$ have not been measured for Ge/SiGe, we expect the order of magnitude to be similar.

\begin{figure}[t]
\centering
\includegraphics[width=0.98\columnwidth]{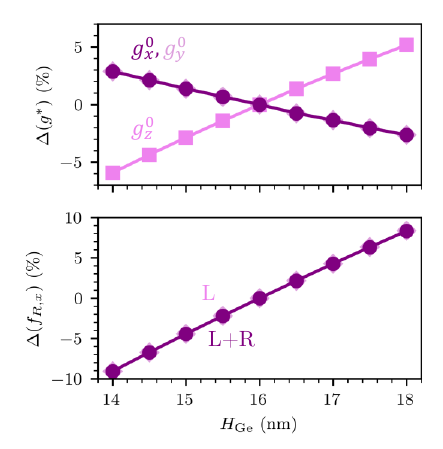}
\caption{a) Dependence of the $g$-factors $g^0_x$, $g^0_y$, $g^0_z$ of the pristine device ($n_i=0$) on the thickness $H_{\rm Ge}$ of the Ge film. The variations $\Delta(g^*)=(g^*-g^*_\mathrm{ref})/g^*_\mathrm{ref}$ are computed with respect to the reference $g^*_\mathrm{ref}$ at $H_{\rm Ge}=16$\,nm. b) Dependence of $f_{R,x}^0$ on $H_{\rm Ge}$ (L and L+R drives). Note that $f_{R,x}^0=0$ for the C drive. $V_{\rm C}$ is adjusted so that $r_\parallel^0=28.3$\,nm for all $H_{\rm Ge}$.}
\label{fig:var_SR}
\end{figure}

In order to quantify the impact of thickness fluctuations on variability, we consider a pristine Ge qubit ($n_i=0$), and compute the dependence of the $g$-factors and Rabi frequencies on the well thickness $H_{\rm Ge}$. The data for $g^0_x$, $g^0_y$, $g^0_z$ and for the Rabi frequency $f^0_{R,x}$ ($\boldsymbol{B}\parallel\boldsymbol{x}$, L and L+R drives) are plotted in Fig. \ref{fig:var_SR}. Note that we did not include the C drive as $f^0_{R,x}=0$ for the pristine device. Also, the charge properties remain almost constant since the radius of the QD is kept fixed ($r_\parallel^0=28.3$\,nm). 

We find that fluctuations of the Ge well thickness by $3\delta z_\mathrm{rms}\approx 2$\,nm induce variations of the in-plane $g$-factors by roughly 3\%. The out-of-plane $g$-factor, which showed remarkably low variability for charge defects, can vary by up to 5\% (due to the strong dependence of $g_z$ on vertical confinement \cite{Michal2021,Martinez2022}). The impact on $f^0_{R,x}$ is even larger, with variations that can reach 10\%. Still, these figures are all smaller than the $\tilde {\mathcal{R}}$'s reported in the main text, so that charge traps usually dominate the overall variability. Long-range fluctuations of $H_{\rm Ge}$ may however become significant in devices where the influence of charge traps has been reduced. This could be achieved, for example, by increasing $H_{\rm SiGe}$ as shown in Appendix \ref{app:thick}, or by engineering the gate layout as discussed in Ref. \cite{Martinez2024}.

\section{Variability in the absence of cool-down strains and in presence of uniaxial strains}\label{app:strains}

Ge hole spins are extremely sensitive to strains. The inhomogeneous strains imprinted by the differential thermal contraction of materials are, notably, a key ingredient for spin manipulation with in-plane magnetic fields \cite{Abadillo2022}. Although strains are not disordered in the present study, they may contribute to the variability of the spin properties as the dot moves in the inhomogeneously deformed crystal under the action of the traps. It is not, therefore, trivial to conclude whether inhomogeneous strains are an asset or a liability. In this Appendix, we discuss variability in the absence of cool-down strains. 

\begin{figure}[t]
\centering
\includegraphics[width=0.98\columnwidth]{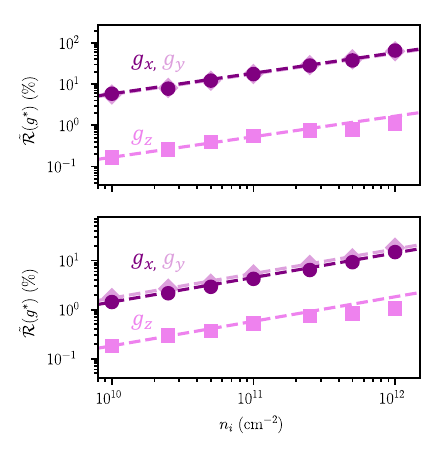}
\caption{a) Relative IQR $\tilde{\mathcal{R}}$ of the $g$-factors $g_x$, $g_y$ and $g_z$ as a function of $n_i$, in the absence of inhomogeneous cool-down strains. b) Same with uniform uniaxial strains $(\varepsilon_{xx}-\varepsilon_{yy})/(2\varepsilon_\parallel)=-10\%$. The calculations are performed at the same $V_{\rm C}$ as in Fig. \ref{fig:charge}.}
\label{fig:var_str}
\end{figure}

The relative IQRs of the effective $g$-factors $g_x$, $g_y$ and $g_z$ are plotted as a function of $n_i$ in Fig. \ref{fig:var_str}. They are similar to those in inhomogeneous strains (see Fig. \ref{fig:g_factors_3}). This shows that the variability of the $g$-factors is mainly due to the disorder-induced deformations of the QDs rather than to their displacements in the inhomogeneous strains. The absolute IQR of $f_{R,x}$ decreases but the relative IQR $\tilde{ \mathcal{R}}(f_{R,x})$ remains, nevertheless, close to 100\% for all drives and $n_i$ (data not shown) because the median Rabi frequencies at $f_L^0=2$\,GHz drop close to 1 MHz. This is in line with the results of Fig. \ref{fig:fR_3}, and the conclusion that inhomogeneous strains are the main mechanism for Rabi oscillations.

A recent work has highlighted the strong impact of uniaxial strains on the gyromagnetic factors of heavy-holes \cite{Mauro2024strain}. The small in-plane $g$-factors can, indeed, be greatly enhanced by the introduction of uniaxial strains $\varepsilon_{xx}\neq\varepsilon_{yy}$. As an illustration, we plot in Fig. \ref{fig:var_str} the relative IQR of $g_x$, $g_y$ and $g_z$ as a function of $n_i$ for homogeneous uniaxial strains $(\varepsilon_{xx}-\varepsilon_{yy})/(2\varepsilon_\parallel)=-10\%$. The $g$-factors of the pristine device (with a uniform charge distribution $n_i=10^{11}$ cm$^{-2}$) are now ${g_x^0}=0.62$ and ${g_y^0}=0.39$. This large increase of the in-plane $g$-factors actually results in a reduction of $\tilde{\mathcal{R}}(g^*_x)$ and $\tilde{\mathcal{R}}(g^*_y)$ by almost a factor 4. Moreover, uniaxial strains efficiently lock the in-plane magnetic axes along $x$ and $y$ (more generally, along the directions parallel and perpendicular to the uniaxial axis). The $\varphi_g$'s are now distributed around 0 with a full width at half-maximum of only $\approx 2^\circ$. This has important consequences for spin manipulation by shuttling \cite{Wang2024}: in the absence of large rotations $\varphi_g$, the required imbalance between the precession axes of the dots is only controlled by the principal $g$-factors $g_x$ and $g_y$; this reduces the probability that neighboring dots have very different random precession axes, but opens the way for a careful control of these axes by the gate voltages. These results show that strain engineering can be an effective strategy to mitigate variability in Ge.

\section{Effect of confinement on variability}\label{app:depVc}

All calculations in the main text are performed at the bias $V_{\rm C}$ such that the radius of the pristine QD is $r_\parallel=r^0_\parallel=28.3$\,nm. This choice is realistic but, nevertheless, arbitrary. In this Appendix, we discuss the dependence of the variability of the charge and spin properties on the confinement strength (i.e., on $V_{\rm C}$) for a trap density $n_i=10^{11}$\,cm$^{-2}$. It is worth reminding that this strength is limited by the band offset between Ge and SiGe, and by the thickness of the top SiGe barrier, as the hole ultimately escapes the quantum well and gets confined at the SiGe/oxide interface at large vertical electric fields. 

\begin{figure}[t!]
\centering
\includegraphics[width=0.98\columnwidth]{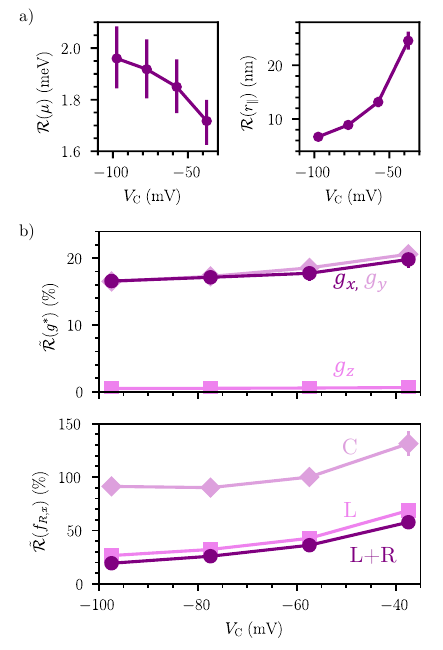}
\caption{Dependence of the variability on $V_{\rm C}$. a) $\mathcal{R}(\mu)$ and $\mathcal{R}(r_\parallel)$ as a function of $V_{\rm C}$ for $n_i=10^{11}$\,cm$^{-2}$. The error bars are the 95\% confidence intervals. b) Relative IQR $\tilde{\mathcal{R}}$ of the $g$-factors $g_x$, $g_y$ and $g_z$, and of $f_R$ (for $\boldsymbol{B}\parallel\boldsymbol{x}$ and the C, L and L+R drives) as a function of $V_{\rm C}$. The data for $g_x$ and $g_y$ are slightly different because the statistics were computed on a smaller set of 500 devices.}
\label{fig:var_vg}
\end{figure}

Fig. \ref{fig:var_vg}a shows the dependence of $\mathcal{R}(\Delta \mu)$ and $\mathcal{R}(r_\parallel)$ on $V_{\rm C}$. While the spread of the QD size is reduced by stronger confinement (more negative $V_{\rm C}$), the IQR of $\Delta\mu$ shows on the opposite a slight increase. Similarly, we plot in Fig. \ref{fig:var_vg}b the relative IQR of $g_x$, $g_y$, and $g_z$ as a function of $V_{\rm C}$, as well as the relative IQR $\tilde{\mathcal{R}}(f_{R,x})$ of the Rabi frequency for the C, L and L+R drives with $\boldsymbol{B}\parallel\boldsymbol{x}$. They only decrease slightly when $V_{\rm C}$ is pulled down. Note that stronger confinement reduces $\mathcal{R}$, but comes along with a decrease of the median $g_x$, $g_y$ and $f_{R,x}$ that smoothes the variations of $\tilde{\mathcal{R}}$. Working with strongly confined QDs can, therefore, homogenize the position and size of the QDs, but does not substantially improve the relative variability of the spin properties.

\section{Dependence of variability on the top SiGe barrier thickness}\label{app:thick}

\begin{figure}[t!]
\centering
\includegraphics[width=0.98\columnwidth]{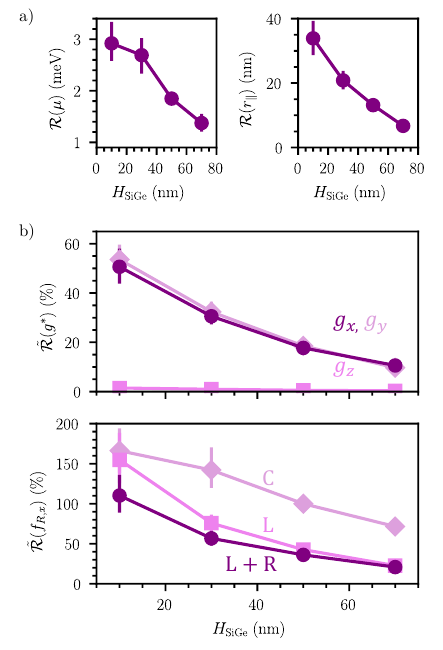}
\caption{Dependence of the variability on the top SiGe barrier thickness. a) $\mathcal{R}(\mu)$ and $\mathcal{R}(r_\parallel)$ as a function of $H_{\rm SiGe}$ for $n_i=10^{11}$\,cm$^{-2}$ and $V_{\rm C}$ adjusted so that $r_\parallel=28.3$\,nm in the pristine QDs. The error bars are the 95\% confidence intervals. b) Relative IQR $\tilde{\mathcal{R}}$ of the $g$-factors $g_x$, $g_y$ and $g_z$, and of $f_R$ (for $\boldsymbol{B}\parallel\boldsymbol{x}$ and the C, L and L+R drives) as a function of $H_{\rm SiGe}$. The data for $g_x$ and $g_y$ are slightly different because the statistics were computed on a smaller set of 500 devices.}
\label{fig:var_diffthick}
\end{figure}

The thickness $H_{\rm SiGe}$ of the top SiGe barrier has a large impact on variability. A thicker barrier indeed shifts the defective interface with the oxide further away from the QDs, at the expense of a weaker electrostatic control (the gates being also farther). As an illustration, we compute the variability due to interface traps with density $n_i=10^{11}$\,cm$^{-2}$ in devices with $H_{\rm SiGe}=10$, 30 and 70\,nm. As in the main text, we set $V_{\rm C}$ so that the radius of the pristine QDs is $r_\parallel=28.3$\,nm. The results are shown in Fig. \ref{fig:var_diffthick}.

With respect to $H_{\rm SiGe}=50$\,nm (main text), reducing the barrier thickness substantially increases the spread of both charge and spin properties. At $H_{\rm SiGe}=10$\,nm, the absolute IQRs of $\mu$ and $d_\parallel$ are $\mathcal{R}(\mu)=2.9$\,meV and $\mathcal{R}(d_\parallel)=34$\,nm, and the relative IQRs or the Larmor and Rabi frequencies are $\tilde{\mathcal{R}}(f_L)\approx 50\%$ and $\tilde{\mathcal{R}}(f_R)\approx 100\%$ (for in-plane magnetic fields). On the opposite, increasing $H_{\rm SiGe}$ beyond 50 nm further helps to mitigate the variability, with $\mathcal{R}(\mu)<1.5$\,meV, $\mathcal{R}(d_\parallel)<10$\,nm, and $\tilde{\mathcal{R}}(f_L,\,f_R)<20\%$ achievable for $H_{\rm SiGe}\gtrsim 70$\,nm. Moreover, the loss of electrostatic control is in fact moderate, as the lever-arm of the C gate only decreases from 0.2 to 0.16, which results in $V_{\rm C}(r_\parallel=28.3\,\textrm{nm})=87$\,mV (as compared to $V_{\rm C}(r_\parallel=28.3\,\textrm{nm})=58$\,mV for $H_{\rm SiGe}=50$\,nm). Consequently, increasing $H_{\rm SiGe}$ can be an effective measure to reduce variability. This conclusion is in line with recent mobility measurements in devices with thick SiGe barriers \cite{Costa2024}. Alternative gate layouts have also been proposed to mitigate variability while preserving a good electrostatic control \cite{Martinez2024}.

\bibliography{var}

@article{Venitucci2018,
  title = {Electrical manipulation of semiconductor spin qubits within the $g$-matrix formalism},
  author = {Venitucci, Benjamin and Bourdet, L\'eo and Pouzada, Daniel and Niquet, Yann-Michel},
  journal = {Physical Review B},
  volume = {98},
  issue = {15},
  pages = {155319},
  numpages = {17},
  year = {2018},
  month = {Oct},
  publisher = {American Physical Society},
  doi = {10.1103/PhysRevB.98.155319},
  url = {https://link.aps.org/doi/10.1103/PhysRevB.98.155319}
}

@article{Michal2021,
  title = {Longitudinal and transverse electric field manipulation of hole spin-orbit qubits in one-dimensional channels},
  author = {Michal, Vincent P. and Venitucci, Benjamin and Niquet, Yann-Michel},
  journal = {Physical Review B},
  volume = {103},
  issue = {4},
  pages = {045305},
  numpages = {17},
  year = {2021},
  month = {Jan},
  publisher = {American Physical Society},
  doi = {10.1103/PhysRevB.103.045305},
  url = {https://link.aps.org/doi/10.1103/PhysRevB.103.045305}}

@article{Maurand2016,
	title = {A {CMOS} Silicon spin qubit},
	volume = {7},
	copyright = {2016 The Author(s)},
	issn = {2041-1723},
	url = {https://www.nature.com/articles/ncomms13575},
	doi = {10.1038/ncomms13575},
	number = {1},
	urldate = {2021-08-31},
	journal = {Nature Communications},
	author = {Maurand, R. and Jehl, X. and Kotekar-Patil, D. and Corna, A. and Bohuslavskyi, H. and Lavi\'{e}ville, R. and Hutin, L. and Barraud, S. and Vinet, M. and Sanquer, M. and De Franceschi, S.},
	month = nov,
	year = {2016},
	pages = {13575},}

@Article{Hendrickx2021,
author={Hendrickx, Nico W. and Lawrie, William I. L. and Russ, Maximilian and van Riggelen, Floor and de Snoo, Sander L. and Schouten, Raymond N. and Sammak, Amir and Scappucci, Giordano and Veldhorst, Menno},
title={A four-qubit Germanium quantum processor},
journal={Nature},
year={2021},
month={Mar},
day={01},
volume={591},
number={7851},
pages={580},
doi={10.1038/s41586-021-03332-6}}

@article{Yang2020,
author = "Yang, C. H. and Leon, R. C. C. and Hwang, J. C. C. and Saraiva, A. and Tanttu, T. and Huang, W. and Camirand Lemyre, J. and Chan, K. W. and Tan, K. Y. and Hudson, F. E. and Itoh, K. M. and Morello, A. and Pioro-Ladri\`{e}re, M. and Laucht, A. and Dzurak, A. S.",
doi = "10.1038/s41586-020-2171-6",
journal = "Nature",
number = "7803",
pages = "350",
title = "Operation of a Silicon quantum processor unit cell above one kelvin",
volume = "580",
year = "2020"
}

@book{Davison1997, 
place = {Cambridge}, 
series = {Cambridge Series in Statistical and Probabilistic Mathematics},
title = {Bootstrap methods and their application}, 
DOI = {10.1017/CBO9780511802843}, 
publisher = {Cambridge University Press}, 
author = {Davison, A. C. and Hinkley, D. V.}, 
year = {1997}, 
collection = {Cambridge Series in Statistical and Probabilistic Mathematics}
}

@article{Martinez2024,
  title = {Mitigating variability in epitaxial-heterostructure-based spin-qubit devices by optimizing gate layout},
  author = {Martinez, Biel and de Franceschi, Silvano and Niquet, Yann-Michel},
  journal = {Phys. Rev. Appl.},
  volume = {22},
  issue = {2},
  pages = {024030},
  numpages = {13},
  year = {2024},
  month = {Aug},
  publisher = {American Physical Society},
  doi = {10.1103/PhysRevApplied.22.024030},
  url = {https://link.aps.org/doi/10.1103/PhysRevApplied.22.024030}
}

@Article{DegliEsposti2024,
author={Degli Esposti, Davide
and Stehouwer, Lucas E. A.
and G{\"u}l, {\"O}nder
and Samkharadze, Nodar
and D{\'e}prez, Corentin
and Meyer, Marcel
and Meijer, Ilja N.
and Tryputen, Larysa
and Karwal, Saurabh
and Botifoll, Marc
and Arbiol, Jordi
and Amitonov, Sergey V.
and Vandersypen, Lieven M. K.
and Sammak, Amir
and Veldhorst, Menno
and Scappucci, Giordano},
title={Low disorder and high valley splitting in silicon},
journal={npj Quantum Information},
year={2024},
month={Mar},
day={13},
volume={10},
number={1},
pages={32},
issn={2056-6387},
doi={10.1038/s41534-024-00826-9},
url={https://doi.org/10.1038/s41534-024-00826-9}
}

@article{Bosco2021,
  title = {Squeezed hole spin qubits in Ge quantum dots with ultrafast gates at low power},
  author = {Bosco, Stefano and Benito, M\'onica and Adelsberger, Christoph and Loss, Daniel},
  journal = {Physical Review B},
  volume = {104},
  issue = {11},
  pages = {115425},
  numpages = {6},
  year = {2021},
  month = {Sep},
  publisher = {American Physical Society},
  doi = {10.1103/PhysRevB.104.115425},
  url = {https://link.aps.org/doi/10.1103/PhysRevB.104.115425}
}

@article{Hanson2007,
  title = {Spins in few-electron quantum dots},
  author = {Hanson, R. and Kouwenhoven, L. P. and Petta, J. R. and Tarucha, S. and Vandersypen, L. M. K.},
  journal = {Rev. Mod. Phys.},
  volume = {79},
  issue = {4},
  pages = {1217},
  numpages = {0},
  year = {2007},
  month = {Oct},
  publisher = {American Physical Society},
  doi = {10.1103/RevModPhys.79.1217},
  url = {https://link.aps.org/doi/10.1103/RevModPhys.79.1217}
}

@Article{Stano2021,
author={Stano, Peter
and Loss, Daniel},
title={Review of performance metrics of spin qubits in gated semiconducting nanostructures},
journal={Nature Reviews Physics},
year={2022},
month={Oct},
day={01},
volume={4},
number={10},
pages={672-688},
issn={2522-5820},
doi={10.1038/s42254-022-00484-w},
url={https://doi.org/10.1038/s42254-022-00484-w}
}

@article{Petta2005,
author = {J. R. Petta  and A. C. Johnson  and J. M. Taylor  and E. A. Laird  and A. Yacoby  and M. D. Lukin  and C. M. Marcus  and M. P. Hanson  and A. C. Gossard },
title = {Coherent manipulation of coupled electron spins in semiconductor quantum dots},
journal = {Science},
volume = {309},
number = {5744},
pages = {2180},
year = {2005},
doi = {10.1126/science.1116955},
URL = {https://www.science.org/doi/abs/10.1126/science.1116955},
not_eprint = {https://www.science.org/doi/pdf/10.1126/science.1116955}}

@Article{Hendrickx2020,
author={Hendrickx, N. W.
and Franke, D. P.
and Sammak, A.
and Scappucci, G.
and Veldhorst, M.},
title={Fast two-qubit logic with holes in Germanium},
journal={Nature},
year={2020},
month={Jan},
day={01},
volume={577},
number={7791},
pages={487},
issn={1476-4687},
doi={10.1038/s41586-019-1919-3},
url={https://doi.org/10.1038/s41586-019-1919-3}
}

@INPROCEEDINGS{DeFranceschi2016,  author={De Franceschi, S. and Hutin, L. and Maurand, R. and Bourdet, L. and Bohuslavskyi, H. and Corna, A. and Kotekar-Patil, D. and Barraud, S. and Jehl, X. and Niquet, Y.-M. and Sanquer, M. and Vinet, M.},  booktitle={2016 IEEE International Electron Devices Meeting (IEDM)},   note={{SOI technology for quantum information processing}},   year={2016},  volume={},  number={},  pages={13.4.1},  doi={10.1109/IEDM.2016.7838409}}

@Article{Piot2022,
author={Piot, N.
and Brun, B.
and Schmitt, V.
and Zihlmann, S.
and Michal, V. P.
and Apra, A.
and Abadillo-Uriel, J. C.
and Jehl, X.
and Bertrand, B.
and Niebojewski, H.
and Hutin, L.
and Vinet, M.
and Urdampilleta, M.
and Meunier, T.
and Niquet, Y.-M.
and Maurand, R.
and Franceschi, S. De},
title={A single hole spin with enhanced coherence in natural silicon},
journal={Nature Nanotechnology},
year={2022},
month={Oct},
day={01},
volume={17},
number={10},
pages={1072-1077},
issn={1748-3395},
doi={10.1038/s41565-022-01196-z},
url={https://doi.org/10.1038/s41565-022-01196-z}
}

@article{Crippa2018,
  title = {Electrical spin driving by $g$-matrix modulation in spin-orbit qubits},
  author = {Crippa, Alessandro and Maurand, Romain and Bourdet, L\'eo and Kotekar-Patil, Dharmraj and Amisse, Anthony and Jehl, Xavier and Sanquer, Marc and Lavi\'eville, Romain and Bohuslavskyi, Heorhii and Hutin, Louis and Barraud, Sylvain and Vinet, Maud and Niquet, Yann-Michel and De Franceschi, Silvano},
  journal = {Physical Review Letters},
  volume = {120},
  issue = {13},
  pages = {137702},
  numpages = {5},
  year = {2018},
  month = {Mar},
  publisher = {American Physical Society},
  doi = {10.1103/PhysRevLett.120.137702},
  url = {https://link.aps.org/doi/10.1103/PhysRevLett.120.137702}
}

@Article{Huang2024,
author={Huang, Jonathan Y.
and Su, Rocky Y.
and Lim, Wee Han
and Feng, MengKe
and van Straaten, Barnaby
and Severin, Brandon
and Gilbert, Will
and Dumoulin Stuyck, Nard
and Tanttu, Tuomo
and Serrano, Santiago
and Cifuentes, Jesus D.
and Hansen, Ingvild
and Seedhouse, Amanda E.
and Vahapoglu, Ensar
and Leon, Ross C. C.
and Abrosimov, Nikolay V.
and Pohl, Hans-Joachim
and Thewalt, Michael L. W.
and Hudson, Fay E.
and Escott, Christopher C.
and Ares, Natalia
and Bartlett, Stephen D.
and Morello, Andrea
and Saraiva, Andre
and Laucht, Arne
and Dzurak, Andrew S.
and Yang, Chih Hwan},
title={{High-fidelity spin qubit operation and algorithmic initialization above 1 K}},
journal={Nature},
year={2024},
month={Mar},
day={01},
volume={627},
number={8005},
pages={772-777},
issn={1476-4687},
doi={10.1038/s41586-024-07160-2},
url={https://doi.org/10.1038/s41586-024-07160-2}
}

@INPROCEEDINGS{Vinet2018,  author={Vinet, M. and Hutin, L. and Bertrand, B. and Barraud, S. and Hartmann, J.-M. and Kim, Y.-J. and Mazzocchi, V. and Amisse, A. and Bohuslavskyi, H. and Bourdet, L. and Crippa, A. and Jehl, X. and Maurand, R. and Niquet, Y.-M. and Sanquer, M. and Venitucci, B. and Jadot, B. and Chanrion, E. and Mortemousque, P.-A. and Spence, C. and Urdampilleta, M. and De Franceschi, S. and Meunier, T.},  booktitle={2018 IEEE International Electron Devices Meeting (IEDM)},   note={{T}owards scalable silicon quantum computing},   year={2018},  volume={},  number={},  pages={6.5.1},  doi={10.1109/IEDM.2018.8614675}}

@Article{Vinet2021,
author={Vinet, Maud},
title={{The path to scalable quantum computing with Silicon spin qubits}},
journal={Nature Nanotechnology},
year={2021},
month={Dec},
day={01},
volume={16},
number={12},
pages={1296},
issn={1748-3395},
doi={10.1038/s41565-021-01037-5},
url={https://doi.org/10.1038/s41565-021-01037-5}
}

@Article{Veldhorst2017,
author={Veldhorst, M.
and Eenink, H. G. J.
and Yang, C. H.
and Dzurak, A. S.},
title={{Silicon CMOS architecture for a spin-based quantum computer}},
journal={Nature Communications},
year={2017},
month={Dec},
day={15},
volume={8},
number={1},
pages={1766},
issn={2041-1723},
doi={10.1038/s41467-017-01905-6},
url={https://doi.org/10.1038/s41467-017-01905-6}
}

@Article{Vandersypen2017,
author={Vandersypen, L. M. K.
and Bluhm, H.
and Clarke, J. S.
and Dzurak, A. S.
and Ishihara, R.
and Morello, A.
and Reilly, D. J.
and Schreiber, L. R.
and Veldhorst, M.},
title={{Interfacing spin qubits in quantum dots and donors -- hot, dense, and coherent}},
journal={npj Quantum Information},
year={2017},
month={Sep},
day={06},
volume={3},
number={1},
pages={34},
issn={2056-6387},
doi={10.1038/s41534-017-0038-y},
url={https://doi.org/10.1038/s41534-017-0038-y}
}

@Article{Mortemousque2021,
author={Mortemousque, Pierre-Andr{\'e}
and Chanrion, Emmanuel
and Jadot, Baptiste
and Flentje, Hanno
and Ludwig, Arne
and Wieck, Andreas D.
and Urdampilleta, Matias
and B{\"a}uerle, Christopher
and Meunier, Tristan},
title={{Coherent control of individual electron spins in a two-dimensional quantum dot array}},
journal={Nature Nanotechnology},
year={2021},
month={Mar},
day={01},
volume={16},
number={3},
pages={296},
issn={1748-3395},
doi={10.1038/s41565-020-00816-w},
url={https://doi.org/10.1038/s41565-020-00816-w}
}

@article{Golovach2006,
  title = {Electric-dipole-induced spin resonance in quantum dots},
  author = {Golovach, Vitaly N. and Borhani, Massoud and Loss, Daniel},
  journal = {Phys. Rev. B},
  volume = {74},
  issue = {16},
  pages = {165319},
  numpages = {10},
  year = {2006},
  month = {Oct},
  publisher = {American Physical Society},
  doi = {10.1103/PhysRevB.74.165319},
  url = {https://link.aps.org/doi/10.1103/PhysRevB.74.165319}
}

@article{Sammak2019,
author = {Sammak, Amir and Sabbagh, Diego and Hendrickx, Nico W. and Lodari, Mario and Paquelet Wuetz, Brian and Tosato, Alberto and Yeoh, LaReine and Bollani, Monica and Virgilio, Michele and Schubert, Markus Andreas and Zaumseil, Peter and Capellini, Giovanni and Veldhorst, Menno and Scappucci, Giordano},
title = {Shallow and undoped Germanium quantum wells: A playground for spin and hybrid quantum technology},
journal = {Advanced Functional Materials},
volume = {29},
number = {14},
pages = {1807613},
doi = {https://doi.org/10.1002/adfm.201807613},
url = {https://onlinelibrary.wiley.com/doi/abs/10.1002/adfm.201807613},
year = {2019}
}

@article{Scappucci2021,
	doi = {10.1557/s43577-021-00147-8},
	url = {https://doi.org/10.1557%2Fs43577-021-00147-8},
	year = 2021,
	month = {jul},
	publisher = {Springer Science and Business Media {LLC}},
	volume = {46},
	number = {7},
	pages = {596},
	author = {G. Scappucci and P. J. Taylor and J. R. Williams and T. Ginley and S. Law},
	title = {Crystalline materials for quantum computing: Semiconductor heterostructures and topological insulators exemplars},
	journal = {{MRS} Bulletin}
}

@Article{Corley-Wiciak2023,
author={Corley-Wiciak, Cedric
and Richter, Carsten
and Zoellner, Marvin H.
and Zaitsev, Ignatii
and Manganelli, Costanza L.
and Zatterin, Edoardo
and Sch{\"u}lli, Tobias U.
and Corley-Wiciak, Agnieszka A.
and Katzer, Jens
and Reichmann, Felix
and Klesse, Wolfgang M.
and Hendrickx, Nico W.
and Sammak, Amir
and Veldhorst, Menno
and Scappucci, Giordano
and Virgilio, Michele
and Capellini, Giovanni},
title={{Nanoscale Mapping of the 3D Strain Tensor in a Germanium Quantum Well Hosting a Functional Spin Qubit Device}},
journal={ACS Applied Materials {\&} Interfaces},
year={2023},
month={Jan},
day={18},
publisher={American Chemical Society},
volume={15},
number={2},
pages={3119-3130},
issn={1944-8244},
doi={10.1021/acsami.2c17395},
url={https://doi.org/10.1021/acsami.2c17395}
}

@Article{Carballido2024,
author={Carballido, Miguel J.
and Svab, Simon
and Eggli, Rafael S.
and Patlatiuk, Taras
and Chevalier Kwon, Pierre
and Schuff, Jonas
and Kaiser, Rahel M.
and Camenzind, Leon C.
and Li, Ang
and Ares, Natalia
and Bakkers, Erik P. A. M.
and Bosco, Stefano
and Egues, J. Carlos
and Loss, Daniel
and Zumb{\"u}hl, Dominik M.},
title={Compromise-free scaling of qubit speed and coherence},
journal={Nature Communications},
year={2025},
month={Aug},
day={15},
volume={16},
number={1},
pages={7616},
issn={2041-1723},
doi={10.1038/s41467-025-62614-z},
url={https://doi.org/10.1038/s41467-025-62614-z}
}

@misc{bassi2024,
      title={{Optimal operation of hole spin qubits}}, 
      author={Marion Bassi and Esteban-Alonso Rodrıguez-Mena and Boris Brun and Simon Zihlmann and Thanh Nguyen and Victor Champain and José Carlos Abadillo-Uriel and Benoit Bertrand and Heimanu Niebojewski and Romain Maurand and Yann-Michel Niquet and Xavier Jehl and Silvano De Franceschi and Vivien Schmitt},
      year={2024},
      eprint={2412.13069},
      archivePrefix={arXiv},
      primaryClass={cond-mat.mes-hall},
      url={https://arxiv.org/abs/2412.13069}, 
}

@Article{Yoneda2018,
author={Yoneda, Jun
and Takeda, Kenta
and Otsuka, Tomohiro
and Nakajima, Takashi
and Delbecq, Matthieu R.
and Allison, Giles
and Honda, Takumu
and Kodera, Tetsuo
and Oda, Shunri
and Hoshi, Yusuke
and Usami, Noritaka
and Itoh, Kohei M.
and Tarucha, Seigo},
title={{A quantum-dot spin qubit with coherence limited by charge noise and fidelity higher than 99.9{\%}}},
journal={Nature Nanotechnology},
year={2018},
month={Feb},
day={01},
volume={13},
number={2},
pages={102},
issn={1748-3395},
doi={10.1038/s41565-017-0014-x},
url={https://doi.org/10.1038/s41565-017-0014-x}
}

@Article{Noiri2022,
author={Noiri, Akito
and Takeda, Kenta
and Nakajima, Takashi
and Kobayashi, Takashi
and Sammak, Amir
and Scappucci, Giordano
and Tarucha, Seigo},
title={{A shuttling-based two-qubit logic gate for linking distant silicon quantum processors}},
journal={Nature Communications},
year={2022},
month={Sep},
day={30},
volume={13},
number={1},
pages={5740},
issn={2041-1723},
doi={10.1038/s41467-022-33453-z},
url={https://doi.org/10.1038/s41467-022-33453-z}
}

@article{Wang2023,
  author = {Wang, Chien-An and Déprez, Corentin and Tidjani, Hanifa and Lawrie, William I. L. and Hendrickx, Nico W. and Sammak, Amir and Scappucci, Giordano and Veldhorst, Menno},
  title = {Probing resonating valence bonds on a programmable germanium quantum simulator},
  journal = {npj Quantum Information},
  year = {2023},
  volume = {9},
  pages = {58},
  doi = {10.1038/s41534-023-00727-3},
}

@Article{Xue2022,
author={Xue, Xiao
and Russ, Maximilian
and Samkharadze, Nodar
and Undseth, Brennan
and Sammak, Amir
and Scappucci, Giordano
and Vandersypen, Lieven M. K.},
title={Quantum logic with spin qubits crossing the surface code threshold},
journal={Nature},
year={2022},
month={Jan},
day={01},
volume={601},
number={7893},
pages={343-347},
issn={1476-4687},
doi={10.1038/s41586-021-04273-w},
url={https://doi.org/10.1038/s41586-021-04273-w}
}

@article{Martinez2022,
  title = {Variability of Electron and Hole Spin Qubits Due to Interface Roughness and Charge Traps},
  author = {Martinez, Biel and Niquet, Yann-Michel},
  journal = {Physical Review Applied},
  volume = {17},
  issue = {2},
  pages = {024022},
  numpages = {24},
  year = {2022},
  month = {Feb},
  publisher = {American Physical Society},
  doi = {10.1103/PhysRevApplied.17.024022},
  url = {https://link.aps.org/doi/10.1103/PhysRevApplied.17.024022}
}

@Article{Borsoi2023,
author={Borsoi, Francesco
and Hendrickx, Nico W.
and John, Valentin
and Meyer, Marcel
and Motz, Sayr
and van Riggelen, Floor
and Sammak, Amir
and de Snoo, Sander L.
and Scappucci, Giordano
and Veldhorst, Menno},
title={Shared control of a 16{\thinspace}semiconductor quantum dot crossbar array},
journal={Nature Nanotechnology},
year={2024},
month={Jan},
day={01},
volume={19},
number={1},
pages={21-27},
issn={1748-3395},
url={https://doi.org/10.1038/s41565-023-01491-3}
}

@Article{Philips2022,
author={Philips, Stephan G. J.
and M{\k{a}}dzik, Mateusz T.
and Amitonov, Sergey V.
and de Snoo, Sander L.
and Russ, Maximilian
and Kalhor, Nima
and Volk, Christian
and Lawrie, William I. L.
and Brousse, Delphine
and Tryputen, Larysa
and Wuetz, Brian Paquelet
and Sammak, Amir
and Veldhorst, Menno
and Scappucci, Giordano
and Vandersypen, Lieven M. K.},
title={Universal control of a six-qubit quantum processor in silicon},
journal={Nature},
year={2022},
month={Sep},
day={01},
volume={609},
number={7929},
pages={919-924},
issn={1476-4687},
doi={10.1038/s41586-022-05117-x},
url={https://doi.org/10.1038/s41586-022-05117-x}
}

@article{Abadillo2022,
  title = {Hole-Spin Driving by Strain-Induced Spin-Orbit Interactions},
  author = {Abadillo-Uriel, Jos\'e Carlos and Rodr\'{\i}guez-Mena, Esteban A. and Martinez, Biel and Niquet, Yann-Michel},
  journal = {Physical Review Letters},
  volume = {131},
  issue = {9},
  pages = {097002},
  numpages = {7},
  year = {2023},
  month = {Sep},
  publisher = {American Physical Society},
  doi = {10.1103/PhysRevLett.131.097002},
}

@article{Rodriguez2025,
  title = {Dressed basis sets for the modeling of exchange interactions in double quantum dots},
  author = {Rodr\'{\i}guez, Mauricio J. and Rodr\'{\i}guez-Mena, Esteban A. and Kalo, Ahmad Fouad and Niquet, Yann-Michel},
  journal = {Phys. Rev. B},
  volume = {112},
  issue = {11},
  pages = {115428},
  numpages = {19},
  year = {2025},
  month = {Sep},
  publisher = {American Physical Society},
  doi = {10.1103/fc5b-5xbx},
  url = {https://link.aps.org/doi/10.1103/fc5b-5xbx}
}

@article{Abadillo2021,
  title = {Two-body Wigner molecularization in asymmetric quantum dot spin qubits},
  author = {Abadillo-Uriel, Jos\'e C. and Martinez, Biel and Filippone, Michele and Niquet, Yann-Michel},
  journal = {Physical Review B},
  volume = {104},
  issue = {19},
  pages = {195305},
  numpages = {17},
  year = {2021},
  month = {Nov},
  publisher = {American Physical Society},
  doi = {10.1103/PhysRevB.104.195305},
  url = {https://link.aps.org/doi/10.1103/PhysRevB.104.195305}
}

@article{Loss1998,
  title = {Quantum computation with quantum dots},
  author = {Loss, Daniel and DiVincenzo, David P.},
  journal = {Physical Review A},
  volume = {57},
  issue = {1},
  pages = {120},
  year = {1998},
  month = {Jan},
  publisher = {American Physical Society},
  doi = {10.1103/PhysRevA.57.120}
}

@article{yu2022strong,
  title={Strong coupling between a photon and a hole spin in silicon},
  author={Yu, C{\'e}cile X and Zihlmann, Simon and Abadillo-Uriel, Jos{\'e} C and Michal, Vincent P and Rambal, Nils and Niebojewski, Heimanu and Bedecarrats, Thomas and Vinet, Maud and Dumur, {\'E}tienne and Filippone, Michele and Bertrand, Beno{\^i}t and De Franceschi, Silvano and Niquet, Yann-Michel and Maurand, Romain},
  journal={Nature Nanotechnology},
  volume={18},
  pages={741},
  year={2023},
  doi = {10.1038/s41565-023-01332-3}
}

@Article{Scappucci2020,
author={Scappucci, Giordano
and Kloeffel, Christoph
and Zwanenburg, Floris A.
and Loss, Daniel
and Myronov, Maksym
and Zhang, Jian-Jun
and De Franceschi, Silvano
and Katsaros, Georgios
and Veldhorst, Menno},
title={The germanium quantum information route},
journal={Nature Reviews Materials},
year={2021},
month={Oct},
day={01},
volume={6},
number={10},
pages={926-943},
issn={2058-8437},
doi={10.1038/s41578-020-00262-z},
url={https://doi.org/10.1038/s41578-020-00262-z}
}

@Article{Jirovec2021,
author={Jirovec, Daniel
and Hofmann, Andrea
and Ballabio, Andrea
and Mutter, Philipp M.
and Tavani, Giulio
and Botifoll, Marc
and Crippa, Alessandro
and Kukucka, Josip
and Sagi, Oliver
and Martins, Frederico
and Saez-Mollejo, Jaime
and Prieto, Ivan
and Borovkov, Maksim
and Arbiol, Jordi
and Chrastina, Daniel
and Isella, Giovanni
and Katsaros, Georgios},
title={A singlet-triplet hole spin qubit in planar Ge},
journal={Nature Materials},
year={2021},
month={Aug},
day={01},
volume={20},
number={8},
pages={1106-1112},
issn={1476-4660},
doi={10.1038/s41563-021-01022-2},
url={https://doi.org/10.1038/s41563-021-01022-2}
}

@Article{Steinacker2024,
author={Steinacker, Paul
and Dumoulin Stuyck, Nard
and Lim, Wee Han
and Tanttu, Tuomo
and Feng, MengKe
and Serrano, Santiago
and Nickl, Andreas
and Candido, Marco
and Cifuentes, Jesus D.
and Vahapoglu, Ensar
and Bartee, Samuel K.
and Hudson, Fay E.
and Chan, Kok Wai
and Kubicek, Stefan
and Jussot, Julien
and Canvel, Yann
and Beyne, Sofie
and Shimura, Yosuke
and Loo, Roger
and Godfrin, Clement
and Raes, Bart
and Baudot, Sylvain
and Wan, Danny
and Laucht, Arne
and Yang, Chih Hwan
and Saraiva, Andre
and Escott, Christopher C.
and De Greve, Kristiaan
and Dzurak, Andrew S.},
title={Industry-compatible silicon spin-qubit unit cells exceeding 99{\%} fidelity},
journal={Nature},
year={2025},
month={Oct},
day={01},
volume={646},
number={8083},
pages={81-87},
issn={1476-4687},
doi={10.1038/s41586-025-09531-9},
url={https://doi.org/10.1038/s41586-025-09531-9}
}

@Article{koch2024,
author={Koch, Thomas
and Godfrin, Clement
and Adam, Viktor
and Ferrero, Julian
and Schroller, Daniel
and Glaeser, Noah
and Kubicek, Stefan
and Li, Ruoyu
and Loo, Roger
and Massar, Shana
and Simion, George
and Wan, Danny
and De Greve, Kristiaan
and Wernsdorfer, Wolfgang},
title={Industrial 300 mm wafer processed spin qubits in natural silicon/silicon-germanium},
journal={npj Quantum Information},
year={2025},
month={Apr},
day={05},
volume={11},
number={1},
pages={59},
issn={2056-6387},
doi={10.1038/s41534-025-01016-x},
url={https://doi.org/10.1038/s41534-025-01016-x}
}

@Article{Hendrickx2023,
author={Hendrickx, N. W.
and Massai, L.
and Mergenthaler, M.
and Schupp, F. J.
and Paredes, S.
and Bedell, S. W.
and Salis, G.
and Fuhrer, A.},
title={Sweet-spot operation of a germanium hole spin qubit with highly anisotropic noise sensitivity},
journal={Nature Materials},
year={2024},
month={Jul},
day={01},
volume={23},
number={7},
pages={920-927},
issn={1476-4660},
doi={10.1038/s41563-024-01857-5},
url={https://doi.org/10.1038/s41563-024-01857-5}
}

@article{Unseld2023,
  author = {Unseld, F. K. and Meyer, M. and Mądzik, M. T. and Borsoi, F. and de Snoo, S. L. and Amitonov, S. V. and Sammak, A. and Scappucci, G. and Veldhorst, M. and Vandersypen, L. M. K.},
  title = "{A 2D quantum dot array in planar 28Si/SiGe}",
  journal = {Applied Physics Letters},
  volume = {123},
  pages = {084002},
  year = {2023},
  doi = {10.1063/5.0160847}
}

@Article{Künne2024,
author={K{\"u}nne, Matthias
and Willmes, Alexander
and Oberl{\"a}nder, Max
and Gorjaew, Christian
and Teske, Julian D.
and Bhardwaj, Harsh
and Beer, Max
and Kammerloher, Eugen
and Otten, Ren{\'e}
and Seidler, Inga
and Xue, Ran
and Schreiber, Lars R.
and Bluhm, Hendrik},
title={The SpinBus architecture for scaling spin qubits with electron shuttling},
journal={Nature Communications},
year={2024},
month={Jun},
day={11},
volume={15},
number={1},
pages={4977},
issn={2041-1723},
doi={10.1038/s41467-024-49182-4},
url={https://doi.org/10.1038/s41467-024-49182-4}
}

@Article{Klemt2023,
author={Klemt, Bernhard
and Elhomsy, Victor
and Nurizzo, Martin
and Hamonic, Pierre
and Martinez, Biel
and Cardoso Paz, Bruna
and Spence, Cameron
and Dartiailh, Matthieu C.
and Jadot, Baptiste
and Chanrion, Emmanuel
and Thiney, Vivien
and Lethiecq, Renan
and Bertrand, Benoit
and Niebojewski, Heimanu
and B{\"a}uerle, Christopher
and Vinet, Maud
and Niquet, Yann-Michel
and Meunier, Tristan
and Urdampilleta, Matias},
title={Electrical manipulation of a single electron spin in CMOS using a micromagnet and spin-valley coupling},
journal={npj Quantum Information},
year={2023},
month={Oct},
day={23},
volume={9},
number={1},
pages={107},
issn={2056-6387},
doi={10.1038/s41534-023-00776-8},
url={https://doi.org/10.1038/s41534-023-00776-8}
}

@article{Takeda2021,
author={Takeda, Kenta
and Noiri, Akito
and Nakajima, Takashi
and Yoneda, Jun
and Kobayashi, Takashi
and Tarucha, Seigo},
title={Quantum tomography of an entangled three-qubit state in silicon},
journal={Nature Nanotechnology},
year={2021},
month={Sep},
volume={16},
pages={965},
doi={10.1038/s41565-021-00925-0}
}

@article{Samkharadze2018,
author = {N. Samkharadze and G. Zheng and N. Kalhor and D. Brousse and A. Sammak and U. C. Mendes and A. Blais and G. Scappucci and L. M. K. Vandersypen},
doi = {10.1126/science.aar4054},
journal = {Science},
pages = {1123},
title = {Strong spin-photon coupling in silicon},
volume = {359},
year = {2018}
}

@Article{HanLim2025,
author={Han Lim, Wee
and Tanttu, Tuomo
and Youn, Tony
and Huang, Jonathan Yue
and Serrano, Santiago
and Dickie, Alexandra
and Yianni, Steve
and Hudson, Fay E.
and Escott, Christopher C.
and Yang, Chih Hwan
and Laucht, Arne
and Saraiva, Andre
and Chan, Kok Wai
and Cifuentes, Jes{\'u}s D.
and Dzurak, Andrew S.},
title={A 2 {\texttimes} 2 Quantum Dot Array in Silicon with Fully Tunable Pairwise Interdot Coupling},
journal={Nano Letters},
year={2025},
month={Jun},
day={16},
publisher={American Chemical Society},
issn={1530-6984},
no_doi={10.1021/acs.nanolett.4c06264},
no_url={https://doi.org/10.1021/acs.nanolett.4c06264}
}

@Article{Zhang2025,
author={Zhang, Xin
and Morozova, Elizaveta
and Rimbach-Russ, Maximilian
and Jirovec, Daniel
and Hsiao, Tzu-Kan
and Fari{\~{n}}a, Pablo Cova
and Wang, Chien-An
and Oosterhout, Stefan D.
and Sammak, Amir
and Scappucci, Giordano
and Veldhorst, Menno
and Vandersypen, Lieven M. K.},
title={Universal control of four singlet--triplet qubits},
journal={Nature Nanotechnology},
year={2025},
month={Feb},
day={01},
volume={20},
number={2},
pages={209-215},
issn={1748-3395},
doi={10.1038/s41565-024-01817-9},
url={https://doi.org/10.1038/s41565-024-01817-9}
}

@misc{tosato2025,
      title={QARPET: A Crossbar Chip for Benchmarking Semiconductor Spin Qubits}, 
      author={Alberto Tosato and Asser Elsayed and Federico Poggiali and Lucas Stehouwer and Davide Costa and Karina Hudson and Davide Degli Esposti and Giordano Scappucci},
      year={2025},
      eprint={2504.05460},
      archivePrefix={arXiv},
      primaryClass={cond-mat.mes-hall},
      url={https://arxiv.org/abs/2504.05460}, 
}

@article{Costa2024,
    author = {Costa, Davide and Stehouwer, Lucas E. A. and Huang, Yi and Martí-Sánchez, Sara and Degli Esposti, Davide and Arbiol, Jordi and Scappucci, Giordano},
    title = {{Reducing disorder in Ge quantum wells by using thick SiGe barriers}},
    journal = {Applied Physics Letters},
    volume = {125},
    number = {22},
    pages = {222104},
    year = {2024},
    month = {11},
    issn = {0003-6951},
    doi = {10.1063/5.0242746},
    url = {https://doi.org/10.1063/5.0242746},
    no_eprint = {https://pubs.aip.org/aip/apl/article-pdf/doi/10.1063/5.0242746/20269738/222104\_1\_5.0242746.pdf},
}

@Article{Seidler2022,
author={Seidler, Inga
and Struck, Tom
and Xue, Ran
and Focke, Niels
and Trellenkamp, Stefan
and Bluhm, Hendrik
and Schreiber, Lars R.},
title={Conveyor-mode single-electron shuttling in Si/SiGe for a scalable quantum computing architecture},
journal={npj Quantum Information},
year={2022},
month={Aug},
day={30},
volume={8},
number={1},
pages={100},
issn={2056-6387},
doi={10.1038/s41534-022-00615-2},
url={https://doi.org/10.1038/s41534-022-00615-2}
}

@Article{vanRiggelen-Doelman2024,
author={van Riggelen-Doelman, Floor
and Wang, Chien-An
and de Snoo, Sander L.
and Lawrie, William I. L.
and Hendrickx, Nico W.
and Rimbach-Russ, Maximilian
and Sammak, Amir
and Scappucci, Giordano
and D{\'e}prez, Corentin
and Veldhorst, Menno},
title={Coherent spin qubit shuttling through germanium quantum dots},
journal={Nature Communications},
year={2024},
month={Jul},
day={08},
volume={15},
number={1},
pages={5716},
issn={2041-1723},
doi={10.1038/s41467-024-49358-y},
url={https://doi.org/10.1038/s41467-024-49358-y}
}

@Article{Dijkema2023,
author={Dijkema, Jurgen
and Xue, Xiao
and Harvey-Collard, Patrick
and Rimbach-Russ, Maximilian
and de Snoo, Sander L.
and Zheng, Guoji
and Sammak, Amir
and Scappucci, Giordano
and Vandersypen, Lieven M. K.},
title={Cavity-mediated iSWAP oscillations between distant spins},
journal={Nature Physics},
year={2025},
month={Jan},
day={01},
volume={21},
number={1},
pages={168-174},
issn={1745-2481},
doi={10.1038/s41567-024-02694-8},
url={https://doi.org/10.1038/s41567-024-02694-8}
}

@Article{Geyer2022,
author={Geyer, Simon
and Het{\'e}nyi, Bence
and Bosco, Stefano
and Camenzind, Leon C.
and Eggli, Rafael S.
and Fuhrer, Andreas
and Loss, Daniel
and Warburton, Richard J.
and Zumb{\"u}hl, Dominik M.
and Kuhlmann, Andreas V.},
title={Anisotropic exchange interaction of two hole-spin qubits},
journal={Nature Physics},
year={2024},
month={Jul},
day={01},
volume={20},
number={7},
pages={1152-1157},
issn={1745-2481},
doi={10.1038/s41567-024-02481-5},
url={https://doi.org/10.1038/s41567-024-02481-5}
}

@Article{Massai2023,
author={Massai, Leonardo
and Het{\'e}nyi, Bence
and Mergenthaler, Matthias
and Schupp, Felix J.
and Sommer, Lisa
and Paredes, Stephan
and Bedell, Stephen W.
and Harvey-Collard, Patrick
and Salis, Gian
and Fuhrer, Andreas
and Hendrickx, Nico W.},
title={Impact of interface traps on charge noise and low-density transport properties in Ge/SiGe heterostructures},
journal={Communications Materials},
year={2024},
month={Aug},
day={14},
volume={5},
number={1},
pages={151},
issn={2662-4443},
doi={10.1038/s43246-024-00563-8},
url={https://doi.org/10.1038/s43246-024-00563-8}
}

@Article{peña2023,
author={Pe{\~{n}}a, Luis Fabi{\'a}n
and Koepke, Justine C.
and Dycus, Joseph Houston
and Mounce, Andrew
and Baczewski, Andrew D.
and Jacobson, N. Tobias
and Bussmann, Ezra},
title={Modeling Si/SiGe quantum dot variability induced by interface disorder reconstructed from multiperspective microscopy},
journal={npj Quantum Information},
year={2024},
month={Mar},
day={27},
volume={10},
number={1},
pages={33},
issn={2056-6387},
doi={10.1038/s41534-024-00827-8},
url={https://doi.org/10.1038/s41534-024-00827-8}
}

@Article{Cifuentes2023,
author={Cifuentes, Jes{\'u}s D.
and Tanttu, Tuomo
and Gilbert, Will
and Huang, Jonathan Y.
and Vahapoglu, Ensar
and Leon, Ross C. C.
and Serrano, Santiago
and Otter, Dennis
and Dunmore, Daniel
and Mai, Philip Y.
and Schlattner, Fr{\'e}d{\'e}ric
and Feng, MengKe
and Itoh, Kohei
and Abrosimov, Nikolay
and Pohl, Hans-Joachim
and Thewalt, Michael
and Laucht, Arne
and Yang, Chih Hwan
and Escott, Christopher C.
and Lim, Wee Han
and Hudson, Fay E.
and Rahman, Rajib
and Dzurak, Andrew S.
and Saraiva, Andre},
title={Bounds to electron spin qubit variability for scalable CMOS architectures},
journal={Nature Communications},
year={2024},
month={May},
day={20},
volume={15},
number={1},
pages={4299},
issn={2041-1723},
doi={10.1038/s41467-024-48557-x},
url={https://doi.org/10.1038/s41467-024-48557-x}
}

@article{wang2024,
author = {Chien-An Wang  and Valentin John  and Hanifa Tidjani  and Cécile X. Yu  and Alexander S. Ivlev  and Corentin Déprez  and Floor van Riggelen-Doelman  and Benjamin D. Woods  and Nico W. Hendrickx  and William I. L. Lawrie  and Lucas E. A. Stehouwer  and Stefan D. Oosterhout  and Amir Sammak  and Mark Friesen  and Giordano Scappucci  and Sander L. de Snoo  and Maximilian Rimbach-Russ  and Francesco Borsoi  and Menno Veldhorst },
title = {Operating semiconductor quantum processors with hopping spins},
journal = {Science},
volume = {385},
number = {6707},
pages = {447-452},
year = {2024},
doi = {10.1126/science.ado5915},
URL = {https://www.science.org/doi/abs/10.1126/science.ado5915}}

@article{Mauro2024,
  title = {Geometry of the dephasing sweet spots of spin-orbit qubits},
  author = {Mauro, Lorenzo and Rodr\'{\i}guez-Mena, Esteban A. and Bassi, Marion and Schmitt, Vivien and Niquet, Yann-Michel},
  journal = {Phys. Rev. B},
  volume = {109},
  issue = {15},
  pages = {155406},
  numpages = {14},
  year = {2024},
  month = {Apr},
  publisher = {American Physical Society},
  doi = {10.1103/PhysRevB.109.155406},
  url = {https://link.aps.org/doi/10.1103/PhysRevB.109.155406}
}

@article{Mauro2024strain,
  title = {Strain engineering in $\mathrm{Ge}$/$\mathrm{Ge}\text{\ensuremath{-}}\mathrm{Si}$ spin-qubit heterostructures},
  author = {Mauro, Lorenzo and Rodr\'{\i}guez-Mena, Esteban A. and Martinez, Biel and Niquet, Yann-Michel},
  journal = {Phys. Rev. Appl.},
  volume = {23},
  issue = {2},
  pages = {024057},
  numpages = {13},
  year = {2025},
  month = {Feb},
  publisher = {American Physical Society},
  doi = {10.1103/PhysRevApplied.23.024057},
  url = {https://link.aps.org/doi/10.1103/PhysRevApplied.23.024057}
}

@article{Martinez2022-inhom,
  title = {Hole spin manipulation in inhomogeneous and nonseparable electric fields},
  author = {Martinez, Biel and Abadillo-Uriel, Jos\'e Carlos and Rodr\'{\i}guez-Mena, Esteban A. and Niquet, Yann-Michel},
  journal = {Phys. Rev. B},
  volume = {106},
  issue = {23},
  pages = {235426},
  numpages = {12},
  year = {2022},
  month = {Dec},
  publisher = {American Physical Society},
  doi = {10.1103/PhysRevB.106.235426},
  url = {https://link.aps.org/doi/10.1103/PhysRevB.106.235426}
}

@article{Reeber96,
author = {Robert R. Reeber and Kai Wang},
doi = {10.1016/S0254-0584(96)01808-1},
journal = {Materials Chemistry and Physics},
number = {2},
pages = {259},
title = {Thermal expansion and lattice parameters of group {IV} semiconductors},
volume = {46},
year = {1996}
}

@Article{Struck2020,
author={Struck, Tom
and Hollmann, Arne
and Schauer, Floyd
and Fedorets, Olexiy
and Schmidbauer, Andreas
and Sawano, Kentarou
and Riemann, Helge
and Abrosimov, Nikolay V.
and Cywi{\'{n}}ski, {\L}ukasz
and Bougeard, Dominique
and Schreiber, Lars R.},
title={Low-frequency spin qubit energy splitting noise in highly purified {28Si/SiGe}},
journal={npj Quantum Information},
year={2020},
month={May},
day={15},
volume={6},
number={1},
pages={40},
issn={2056-6387},
doi={10.1038/s41534-020-0276-2},
url={https://doi.org/10.1038/s41534-020-0276-2}
}

@article{cvitkovich2024,
  title = {Coherence limit due to hyperfine interaction with nuclei in the barrier material of $\mathrm{Si}$ spin qubits},
  author = {Cvitkovich, Lukas and Stano, Peter and Wilhelmer, Christoph and Waldh\"or, Dominic and Loss, Daniel and Niquet, Yann-Michel and Grasser, Tibor},
  journal = {Phys. Rev. Appl.},
  volume = {22},
  issue = {6},
  pages = {064089},
  numpages = {16},
  year = {2024},
  month = {Dec},
  publisher = {American Physical Society},
  doi = {10.1103/PhysRevApplied.22.064089},
  url = {https://link.aps.org/doi/10.1103/PhysRevApplied.22.064089}
}

@article{Bauza02,
author = {Bauza, D.},
journal = {IEEE Electron Device Letters}, 
title = {Extraction of {Si-SiO$_2$} interface trap densities in {MOS} structures with ultrathin oxides}, 
year = {2002},
volume = {23},
number = {11},
pages = {658},
doi = {10.1109/LED.2002.805008}
}

@INPROCEEDINGS{Brunet09,
author = {Brunet, L. and Garros, X. and Andrieu, F. and Reimbold, G. and Vincent, E. and Bravaix, A. and Boulanger, F.},
booktitle = {2009 IEEE International SOI Conference}, 
title = {New method to extract interface states density at the back and the front gate interfaces of {FDSOI} transistors from {CV-GV} measurements}, 
year = {2009},
volume = {},
number = {},
pages = {1},
doi = {10.1109/SOI.2009.5318747}
}

@article{Pirro16,
author = {Pirro, L. and Ionica, I. and Ghibaudo, G.  and Mescot, X. and Faraone, L. and Cristoloveanu, S.},
title = {Interface trap density evaluation on bare silicon-on-insulator wafers using the quasi-static capacitance technique},
journal = {Journal of Applied Physics},
volume = {119},
number = {17},
pages = {175702},
year = {2016},
doi = {10.1063/1.4947498},
}

@Article{john2024,
author={John, Valentin
and Yu, C{\'e}cile X.
and van Straaten, Barnaby
and Rodr{\'i}guez-Mena, Esteban A.
and Rodr{\'i}guez, Mauricio
and Oosterhout, Stefan D.
and Stehouwer, Lucas E. A.
and Scappucci, Giordano
and Rimbach-Russ, Maximilian
and Bosco, Stefano
and Borsoi, Francesco
and Niquet, Yann-Michel
and Veldhorst, Menno},
title={Robust and localised control of a 10-spin qubit array in germanium},
journal={Nature Communications},
year={2025},
month={Nov},
day={26},
volume={16},
number={1},
pages={10560},
issn={2041-1723},
doi={10.1038/s41467-025-65577-3},
url={https://doi.org/10.1038/s41467-025-65577-3}
}

@article{Vermeer21,
author={Vermeer, M. L. and Hueting, R. J. E. and Pirro, L. and Hoentschel, J. and Schmitz, J.},
journal={IEEE Transactions on Electron Devices}, 
title={Interface States Characterization of {UTB} {SOI} {MOSFETs} From the Subthreshold Current}, 
year={2021},
volume={68},
number={2},
pages={497},
doi={10.1109/TED.2020.3043223}
}

@article{Terrazos21,
title = {Theory of hole-spin qubits in strained germanium quantum dots},
author = {Terrazos, L. A. and Marcellina, E. and Wang, Zhanning and Coppersmith, S. N. and Friesen, Mark and Hamilton, A. R. and Hu, Xuedong and Koiller, Belita and Saraiva, A. L. and Culcer, Dimitrie and Capaz, Rodrigo B.},
journal = {Physical Review B},
volume = {103},
pages = {125201},
year = {2021},
doi = {10.1103/PhysRevB.103.125201},
}

@article{Marcellina17,
title = {Spin-orbit interactions in inversion-asymmetric two-dimensional hole systems: A variational analysis},
author = {Marcellina, E. and Hamilton, A. R. and Winkler, R. and Culcer, Dimitrie},
journal = {Physical Review B},
volume = {95},
pages = {075305},
year = {2017},
doi = {10.1103/PhysRevB.95.075305},
}

\end{document}